\documentclass[12pt,preprint]{aastex}

\newcommand{\za}{$z_{\rm abs}$}
\newcommand{\zc}{$z_{\rm cluster}$}

\newcommand{\zave}{$\langle z \rangle$}
\newcommand{\nh}{$N_{\rm hit}$}

\newcommand{\hi}{\ion{H}{1}}
\newcommand{\mgii}{\ion{Mg}{2}}
\newcommand{\civ}{\ion{C}{4}}
\newcommand{\ovi}{\ion{O}{6}}

\newcommand{\icm}{cm$^{-2}$}
\newcommand{\kms}{km~s$^{-1}$}
\newcommand{\lya}{Ly$\alpha$\ }

\newcommand{\hkpc}{$h^{-1}_{71}$ kpc}
\newcommand{\hmpc}{$h^{-1}_{71}$ Mpc}

\slugcomment{Accepted for publication in ApJ}

\shorttitle{\mgii\ in Cluster Galaxies}
\shortauthors{Lopez et al.}

\begin{document}


\title{Galaxy Clusters in the Line of Sight to Background Quasars: \\ I. Survey
  Design and Incidence of \mgii\ Absorbers at
  Cluster 
  Redshifts\footnote{This paper includes data gathered with the 6.5 meter
  Magellan Telescopes located at Las Campanas Observatory, Chile.}  
}



\author{S. Lopez\altaffilmark{1}, 
L. F. Barrientos\altaffilmark{2},
P. Lira\altaffilmark{1}, 
N. Padilla\altaffilmark{2},
D. G. Gilbank\altaffilmark{3},
M. D. Gladders\altaffilmark{4,}\altaffilmark{5},
J. Maza\altaffilmark{1},
N. Tejos\altaffilmark{1},
M. Vidal\altaffilmark{1}, \& 
H. K. C. Yee\altaffilmark{3}
}






\altaffiltext{1}{Departamento de Astronom\'ia, Universidad de Chile, Casilla
     36-D, Santiago, Chile.}
\altaffiltext{2}{Departamento de Astronom\'ia y Astrof\'isica, Universidad
     Cat\'olica de Chile, Avenida Vicu\~a Mackenna 4860, Casilla 306, Santiago
     22, Chile.} 
\altaffiltext{3}{Department of Astronomy and Astrophysics, University of
     Toronto, 50 St. George Street, Toronto, ON M5S 3H4, Canada.}  
\altaffiltext{4}{Department of Astronomy and Astrophysics, University of
     Chicago, 5640 South Ellis Avenue, Chicago, IL 60637, USA.} 
\altaffiltext{5}{Visiting Associate, The
  Observatories of the Carnegie Institution of Washington, 813 Santa
  Barbara St., Pasadena, CA 91101, USA.}



\begin{abstract}
Quasar absorption line systems are redshift-independent sensitive mass
tracers. Here we describe the first optical survey of absorption
systems associated with galaxy clusters at $z= 0.3-0.9$. We have
cross-correlated quasars from the third data release of the Sloan
Digital Sky Survey with high-redshift cluster/group candidates from
the Red-Sequence Cluster Survey. In a common field of $\approx 20$
square degrees, we have found $442$ quasar-cluster pairs for which the
\mgii\ $\lambda\lambda 2796,2803$ \AA\ doublet might be detected at a
transverse (physical) distance $< 2$ \hmpc\ from the cluster
centers. In addition, we have found 33 other pairs in the literature
and we have discovered 7 new quasars with foreground clusters.  To
investigate the incidence $(dN/dz)$ and equivalent-width distribution
$n(W)$ of \mgii\ systems at cluster redshifts, two
statistical samples were drawn out of these pairs: one made of
high-resolution spectroscopic quasar observations (46 pairs), and one
made of quasars used in \mgii\ searches found in the literature (375
pairs).  The total redshift path from an ad-hoc definition of 'cluster
redshift path' is $\Delta$\zc\ $=6.3$ and $\Delta$\zc\ $=57.0$ for the
two samples, respectively. We estimate the completeness level to be
nearly 100 \%\ for $W$ detection thresholds of
$W_0^{2796}>0.05$ and $W_0^{2796}>1.0$ \AA\ in the two samples,
respectively.

The results are: 
(1) 
the population of {\it strong}  \mgii\ systems
($W_0^{2796}>2.0$ \AA) near cluster redshifts shows a significant ($>3\sigma$)
overabundance (up to a factor of $15$) when compared with the 'field'
population;  
(2) 
the overabundance is more evident at smaller
distances ($d<1$ \hmpc) than larger distances ($d<2$ \hmpc) from the cluster
center; and, 
(3) 
the population of {\it weak} \mgii\ systems
($W_0^{2796}<0.3$ \AA) near cluster redshifts conform to the field
statistics.
Unlike in the field, this dichotomy makes $n(W)$ in clusters appear flat  
and well fitted by a power-law in the entire $W$-range.  We assess
carefully all possible selection and systematic effects, and conclude that the
signal is indeed due to the presence of clusters. In particular, a sub-sample
of the most massive clusters yields a stronger and still significant signal.
Since either the absorber number density or filling-factor/cross-section
affects the absorber statistics, an interesting possibility is that we have
detected the signature of truncated halos due to environmental effects. Thus,
we argue that the excess of strong systems is due to a population of absorbers
in an overdense galaxy environment, and the lack of weak systems to a
different population, that got destroyed in the cluster environment. 

Finally, comparison with models of galaxy counts  show that there is
proportionally less cold gas in more massive clusters than in low-mass
systems, and two orders of magnitude less \mgii\ cross-section due to weak
systems than due to stronger systems.
\end{abstract}



\section{Introduction}

Galaxy clusters trace the densest environments in the Universe. They thus
  constitute the best laboratories to study galaxy evolution since (1) they
  contain a large number of galaxies at essentially the same cosmic time, (2)
  their environment is extreme compared to the field so galaxy transformations
  are constantly present, and (3) they can be traced to large lookback
  times. Yet the baryon budget in clusters is not all that well constrained
  mainly because it is not clear whether all baryonic constituents have been
  identified and quantified (e.g., Ettori 2003; McCarthy, Bower, \& Balogh
  2007). According to Ettori (2003) these constituents are: hot baryons
  (intracluster medium, 70\%), cold baryons (galaxies, stars and gas, 13\%),
  and warm baryons (unknown, 17 \%).

In addition to detecting galaxies and the intracluster medium in emission,
clusters have recently been probed through absorption by metals in x-ray
spectra of background AGNs (Takei et al. 2007). However, this absorption is
rather associated with the hot intracluster gas and not with the cluster
galaxies. Since gas associated with {\it field} galaxies is known to produce
detectable EUV absorption in background quasar spectra, one could in principle
probe the {\it cold-warm} ($T<10^5$ K) gas associated with cluster galaxies
using this quasar absorption line (QAL) technique. One great advantage of the
QAL technique is that it provides a sensitive measure of the gas that is
independent of redshift and host-galaxy brightness.

In this paper we present the first spectroscopic survey of background
quasars having foreground clusters in the line of sight. The survey is
aimed at probing metal absorbers possibly associated with the cluster
galaxies. We concentrate on the incidence of the redshifted \mgii\
$\lambda\lambda 2796,2803$ \AA\ doublet, an excellent tracer of
high-redshift galaxies (Bergeron \& Stasinska 1986; Petitjean \&
Bergeron 1990; Steidel \& Sargent  2002; Churchill et al. 2000; Zibetti
et al. 2007) for which extensive field surveys exist. The \mgii\
doublet has been used extensively in spectroscopic quasar surveys
because it is a strong and an easy-to-find transition, and has a
redshift coverage from the ground of \za$\ga 0.2$, matching imaging
studies.  Redshifted metal absorption lines in a quasar spectrum
appear together with absorption by neutral hydrogen  in what is called
'absorption systems'. The incidence of absorption systems, $dN/dz$,
i.e., the probability of line-of-sight (LOS) intersection per unit
redshift, and its equivalent width distribution, $\frac{d^2N}{dWdz}$,
are important observables as they depend both on the absorbing
cross-section and number density of the absorbers. More importantly,
these quantities can be measured without previous knowledge of the
nature and environment of the absorbers, i.e., galaxies, \lya\ forest,
etc.

Early \mgii\ surveys (e.g., Lanzetta, Turnshek, \& Wolfe 1987, Tytler
et al. 1987, Steidel \& Sargent 1992), sensitive to a rest-frame
equivalent width (rEW) threshold of $W_0^{2796}>0.3$ \AA, established
a population of non-evolving absorbers up to $z=2$ with signs of
clustering on scales $< 500$ \kms\ (Petitjean \& Bergeron 1990;
Steidel \& Sargent 1992). More recent surveys (Churchill et al. 1999
[hereafter CRCV99]; Nestor, Turnshek \& Rao. 2005 [NTR05]; Nestor,
Turnshek \& Rao 2006 [NTR06], Narayanan et al. 2007, Lynch, Charlton
\& Kim 2006; Prochter, Prochaska, \& Burles 2006 [PPB06]) have shown a
clear dichotomy between strong and weak absorbers: the equivalent
width distribution is steeper for weak systems than for strong ones,
with a transition around $W_0^{2796}\approx0.3$ \AA. This has led some
authors to propose different populations/environments for these two
classes of systems (NTR06).

On the other hand, surveys of {\it galaxies} selected by \mgii-absorption have
shown a population of normal morphology, bright galaxies, with absorption
cross sections that range from a few to several tens of \hkpc\ depending on
rEW.  \mgii\ was linked to bright galaxies early in the 90's thanks to the
work by Steidel \& Sargent (1992), Bergeron \& Boiss\'e (1991), Lanzetta \&
Bowen (1990), Le Brun et al. (1993), among others, and more recently to
rotating disks (Steidel et al. 2002),
neutral gas (Ellison et al. 2004a; Rao, Turnshek \& Nestor 2006), and
also to large-scale structure (Williger et al. 2002).  Although it seems clear
that \mgii\ absorption arises in galaxies of a wide range of morphologies and
luminosities (Kacprzak et al. 2007), the majority of the strong systems could
be associated with blue, starburst galaxies (Zibetti et al. 2007) with high
metallicties (Ellison, Kewley, \& Mall\'en-Ornelas 2005). However, none of
these identifications tells us where and through which processes the
absorption occurs in these galaxies.  If the \mgii\ occurs in extended
  halos, the covering factor may be less than unity, so the halos must be
  patchy (Churchill et al. 2005; Churchill et al. 2007). Indeed, this
``patchiness'' may point out to alternative explanations like \mgii\ systems
being the high-redshift analogs of local HVCs; i.e., warm ($10^4$ K), massive
($10^6$ M$_\sun$) and compact, pressure-confined clouds embedded in a hot halo
but still virialized (e.g., Maller \& Bullock 2004), or, alternatively, part
of cool galactic outflows (Bouche et al. 2006 [BMPCW06]). In any case, and
despite a yet unclear origin, there is overall consensus that \mgii\ flags
star-forming regions in a variety of galaxies.

Do {\it cluster} galaxies host \mgii\ absorbers? This question
motivates the present paper.  Cluster galaxy properties are
essentially different from field galaxies due to environmental
effects. While the general galaxy population shows a wide range of
mass, morphology, gas and stellar content, and halo sizes, some of
these properties have been found to depend strongly on their local
galaxy density. For instance, in the morphology-density relation
(Dressler 1980) early-type galaxies are concentrated toward the cores
of the galaxy clusters, while late-type galaxies are found mainly in
the lower density environments ('cluster suburbs' or the
'field'). Similarly, the increasing fraction of blue galaxies in
clusters with increasing redshift --the Butcher-Oemler effect (Butcher
and Oemler 1984)-- was the first indication that the population of
galaxies evolved. 

Thus, clearly, detecting and studying \mgii\ absorption in overdense regions
like cluster galaxies has a twofold potential. It provides constraints to
fundamental field properties of the absorption systems (clustering, halo
masses, and the absorber-IGM connection); on the other hand, it also provides
independent clues to galaxy accretion and evolution in clusters,  which may
become a key complement to radio observations of cold gas in local and
low-redshift cluster galaxies (Chung et al. 2007; Vollmer et al. 2007;
Verheijen et al. 2007).




Our  paper is organized as follows: we first describe the quasar-cluster
correlations in \S~\ref{sect-pairs}, then we describe the spectroscopic quasar
observations in \S~\ref{sect-obs}. In \S~\ref{sect-sample} we define the
samples and explain the method to get the \mgii\ statistics in clusters, while
in \S~\ref{sect-results} we present the results. An assessment of survey
completeness and biases is presented in \S~\ref{sect-stats}. Finally, we
summarize the results in \S~\ref{sect-summary} and discuss the implications in
\S~\ref{sect-discussion}. Throughout the paper we use a cosmology with
$(\Omega_M,\Omega_\Lambda)=(0.27,0.73)$ and $H_0\equiv 71~h_{71}$
km~s$^{-1}$/Mpc.

\section{Selection of quasar-cluster pairs\label{sect-pairs}}

Our primary goal is to study the incidence of \mgii\ absorbers in galaxies
associated with cluster galaxies, and this over an as wide as possible
range of line strengths. To this aim a sample must be built that includes
bright quasars (suitable for high-resolution spectroscopy) close in projection
to and at higher redshifts than the clusters.

We searched for potential quasar-cluster pairs in three ways: (1) search
for known Sloan Digital Sky Survey (SDSS) quasars in fields of cluster
candidates from the Red-Sequence Cluster Survey (RCS); (2) search for known or
new quasars in fields of clusters from the Chandra database; (3) search
in the NASA/IPAC Extragalactic Database (NED) for quasars 
close in projection to objects labeled as clusters or 
groups.

In the search we have imposed two broad criteria\footnote{Further, tighter
  criteria are applied later when we describe the statistical samples in
  \S~\ref{sect-stat}}: (1) for each quasar-cluster pair we require $0.2\le
  z_{\rm cluster}\le z_{\rm quasar}$, i.e., the redshifted \mgii\ doublet may
  be detected at the cluster redshift, \zc, and is observable from the ground;
  and (2) at \zc\ the quasar line of sight (LOS) lies within a transverse
  (physical) distance of $d=2$ \hmpc\ of cluster coordinates (this distance
  was considered enough to probe well beyond the virial radius).  We will
  refer to these criteria as the ``quasar-cluster'' criteria.

\subsection{Cross-correlation of SDSS quasars and RCS
  clusters\label{sect-correlation}} 

We describe the cross-correlation of cluster candidates from the RCS with
quasars from the SDSS data release three (SDSS-DR3; Schneider et al. 2005). We
do not use a later release because most of the extant \mgii\ statistics were
obtained using DR3 data.
  
The RCS (Gladders \& Yee 2005) is a $\sim 100$ square degree optical
survey conducted at CFHT 
and CTIO, aimed at finding galaxy clusters up to redshift of one with
some sensitivity to massive clusters to $z\sim1.4$. This survey has
been carried out with observations in two bands, $R$ and $z'$, to
obtain galaxy colors and thus to enhance the contrast between cluster
and field galaxies (Gladders \& Yee 2000). The main goal of the survey
is to measure cosmological parameters through the evolution of the
cluster mass function (Gladders et al. 2007).

The clusters have been selected from an overdensity in position, color and
magnitude, and their redshifts have been determined from the loci of the
red-sequence in the color-magnitude diagram. The redshifts were estimated from
Simple Stellar Population codes and then calibrated through the comparison
with spectroscopic redshifts for a sample at a wide range of redshifts
(Gilbank et al. 2007). Masses for the different clusters were determined by
using the optical richness measured by the $B_{gc}$ parameter (Yee \&
Ellingson 2003), and the relationship between $B_{gc}$ and $M_{200}$  (the
  mass interior to $r_{200}$, where the average mass density is $200\rho_c$)
  in Yee \& Ellingson (2003; see also Gladders et al. 2007).  Spectroscopy
shows that the contamination of the RCS cluster sample, even at $z\sim 1$, is
less than 10\% (Gilbank et al. 2007; Barrientos et al., in prep.), and as low
as 3\% at lower redshifts (Blindert et al., in prep.).

Note that the RCS cluster sample we use is an inclusive sample of all
RCS cluster candidates with no redshift restrictions (other than the
natural ones imposed by the survey design) and no richness cuts. Thus,
it is likely less clean than the restricted best sample used in the
analysis of Gladders et al. (2007); however, the inclusion of all
candidates maximises possible overlap with the SDSS quasar sample.

Although covering basically different areas, the cross-correlation of RCS
clusters with SDSS quasars from DR3 yielded 442 quasar-cluster pairs that met
the quasar-cluster criteria (113 for $d< 1$ \hmpc\ and 36 for $d<0.5$
\hmpc). These quasar-cluster pairs are distributed in a common area of
$\approx 20$ square degrees. We will refer to this sample as the ``SDSS-RCS
sample''. This sample contributes the vast majority of pairs used in the
present study. Later in this paper we select quasar-cluster pairs from a
sub-sample of rich clusters.   

\begin{figure}
\epsscale{.80}
\includegraphics[angle=-90,scale=.8]{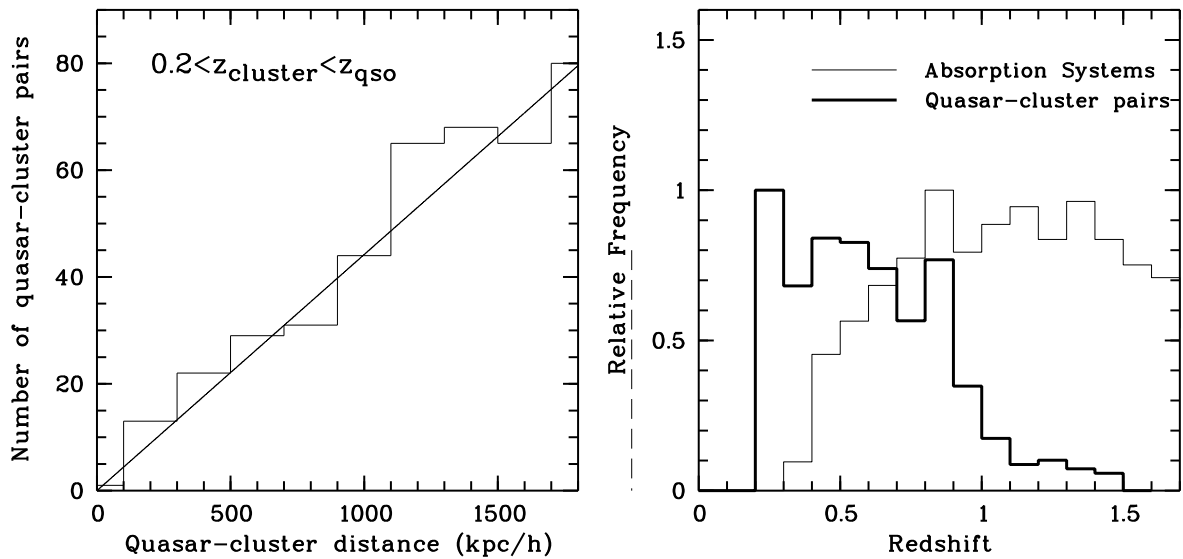}
\caption{ {\it Left:} Number of quasar-cluster pairs in the SDSS-RCS
  sample as a function of the projected physical distance between
  cluster and quasar line-of-sight at cluster redshift. The line is
  the expectation for constant projected number density of pairs. {\it
  Right:} Redshift distribution (normalized to maximum frequency) of
  clusters in the SDSS-RCS sample and of \mgii\ absorbers in Prochter,
  Prochaska \& Burles (2006).\label{fig_histo}}
\end{figure}

Fig.~\ref{fig_histo} shows the transverse-distance and
cluster-redshift distributions of the SDSS-RCS sample. In the first
one we plot the number of pairs found to have a given quasar-cluster
distance.  To see that this distribution results from a random
distribution of clusters and quasars, we calculate the expected
distribution (the stright line in the Figure) defining a mean density
as the total number of pairs divided by the area of a circle of radius
$2\,000$ \hkpc. This comparison shows that all distances are well
represented and that they roughly follow a uniform distribution, which
is important for the homogeneity of the survey. The righthand panel of
the Figure shows that the redshift distribution of \mgii\ systems
found in SDSS quasar spectra (thin line; PPB06) and that of the
SDSS-RCS sample have considerable overlap, meaning that our
cross-correlation is well suited for searches of \mgii\ in cluster
galaxies.

The SDSS-RCS sample of 442 quasar-cluster pairs is composed of 190 quasars and
368 clusters. Therefore, there are on average $\approx 2$ clusters per LOS, and
$\approx 20$ \% of the clusters are crossed by more than one LOS.  Regarding
observability, roughly $80$ \%\ of the quasars are brighter than $g=20$ mag,
and $\approx 50$ \% of them are observable from Southern facilities.

\subsection{New x-ray selected quasars}

Since both galaxy clusters and quasars are ubiquitous x-ray emitters, 
using archival Chandra observations proved to be a successful way of
selecting further targets for our study.

From the Chandra database we selected all public observations under
the science category `Galaxy Clusters'. We imposed a maximum
declination of $+20$ degrees, and a Chandra exposure time $\ga 25$
ksec to ensure significant detections of the quasar candidates. The
clusters also had to have a determined redshift above $z=0.2$. A
final list of 29 observations that met these criteria were retrieved
from the archive.

Next, we identified point-like sources in the x-ray data. We looked for
candidate quasars located within a radius of $\la 5\arcmin$ from the cluster
central position.  Since the observations were aimed at the cluster centers,
we did not have to worry about the degradation of the Chandra point spread
function with increasing off-axis distances. We then searched for optical
point-like counterparts to the x-ray sources in SDSS images and obtained their
$R$ and $B$ magnitudes from the APM catalog. Imposing the criteria $R > 16$
and $B-R \la 2.0$, a total of 49 candidate quasars were selected in 23 of the
Chandra clusters.  We will refer to this sample as the ``x-ray sample''.

\subsection{Pairs from the literature}

A search in the NED was performed of quasars near RCS coordinates. Out
of 7263 searches, 28 yielded quasars not found by the SDSS, that met
the quasar-cluster criteria.  We will refer to this sample as the
``literature sample''. In addition, 5 other quasar-cluster pairs found
in the literature were added to this sample. It is important to note
that SDSS clusters are not well suited for our study due to their
lower redshift ($z<0.3$; Koester et al. 2007).

\section{Observations\label{sect-obs}}

\subsection{Low-resolution spectroscopy\label{low-res}}

Low resolution optical spectroscopic follow-up observations of the
quasar candidates from the x-ray sample were carried out with the Wide
Field Reimaging CCD Camera in long-slit grism mode on the du Pont
telescope at Las Campanas Observatory on March 30 and September 15-16,
2006. We used the blue grism, which gives a resolution of $\sim 3$\AA\
and a wavelength range of $\sim 4700$\AA.

Sixteen candidates were observed with enough signal-to-noise ratio to determine
emission redshifts, and out of these, 7 quasars were confirmed. Other
counterparts corresponded to Seyfert and star-forming galaxies, and a few
stars (probably due to chance alignments). Therefore, the technique of using
x-ray data to find quasars gave a success rate of $\approx 45$ \%.

\subsection{High-resolution spectroscopy}

Echelle spectra were obtained using the MIKE spectrograph on the Las Campanas
Clay 6.5m telescope. We obtained 18 quasar spectra in three runs on March
18-19 and September 23-24 and 29-30, 2006. Twelve of the quasars are from the
SDSS-RCS, 2 from the x-ray, and 4 from the literature samples.   The
  target selection was based only on airmass and brightness, i.e., without
  consideration of cluster redshifts. The observed sample represents $\approx
  15$ \%\ of the total number of available targets in the three samples.

Weather conditions were good but quite variable for two of the runs. Seeing
ranged from good ($1\arcsec$) to excellent ($0.6\arcsec$).  We made best
efforts to obtain a S/N ratio as homogeneous as possible throughout the
sample.

MIKE is mounted on the Nasmyth port and the slit orientation on the
plane of the sky is fixed. For long exposures, and despite a low
airmass, this requires manual corrections to keep the object centered
on the slit, a task that proved feasible in general but difficult to
carry out for some mag $\approx 20$ targets. For five of our targets
we used integration times in excess of 4 hours. All spectra were taken
with a $1\arcsec$ slit and an on-chip binning of $2\times3$ pixels.
With this setup the final spectral resolution of our spectra was
$\approx12.0$ and $\approx 13.5$ \kms\ (FWHM) for the blue and red arms,
respectively.

To extract the spectra we used our own pipeline running on MIDAS. The
two-dimensional echelle spectra were flat-fielded (using star spectra taken
with a diffusor) and extracted optimally (fitting the seeing profiles and
taking into account the spatial tilts introduced by the cross-dispersing
prisms). The orders were then calibrated with spectra of a Thorium-Argon lamp
(using typically 15-20 lines per echelle order) and the different exposures
co-added using a vacuum-heliocentric scale with $\Delta\lambda=0.067565$ and
$0.1447107$ \AA\ for the blue and red orders, respectively.  Finally, the
orders were normalized and merged. The spectral coverage of each spectrum is
$\lambda=3\,350$ to $7\,480$ \AA. Table~\ref{tbl-obs} summarizes the
echelle  observations.

%
%
%

\section{Sample Definitions and Redshift Path
  Density\label{sect-stat}\label{sect-sample}}

In what follows we describe the various statistical samples drawn from the
data. These samples were derived from the data on absorbers (see
Table~\ref{tbl-qsos}) and clusters (Table~\ref{tbl-clusters}).  We define the
'cluster redshift-path' of the survey and the sample of 'hits', or absorption
systems found in the cluster redshift-path (summarized in
Tables~~\ref{tbl-samples} and~\ref{tbl-hits}).

\subsection{Sample of \mgii\ Absorption Systems}

\subsubsection{\mgii\ in High-resolution Spectra (Sample 'S1')}

The 18 MIKE spectra along with one UVES spectrum from the Literature Sample
define what we shall call the 'high-resolution sample', hereafter '{\it S1}'.
As in previous high-resolution surveys (e.g., Narayanan et al. 2007), we searched visually for \mgii\ systems in {\it S1} by carefully scanning
redshift chunks all along the range of \mgii\ detectability, each time
plotting in velocity both doublet lines. We considered lines detected at the
$3\sigma$ level or higher in {\it both} doublet lines.

Table~\ref{tbl-qsos} presents the absorption line data (LOS up to entry 19
in  {\it S1}). Absorption redshifts are determined to within $\delta
z_{\rm abs} \sim 10^{-4}$.  rEW were calculated using pixel
integrations with $1\sigma$ errors from propagated pixel variances.  Lines
within a velocity window of $500$ \kms\ were considered one system, to conform
to previous QAL surveys. Column '$z_{\rm EW}$' displays the minimum redshift
at which a line with $W_0=0.05$ \AA\ can be detected at the $3\sigma$
significance level. This value was computed assuming the error in the observed
rEW is given by $\sigma_W={\rm FWHM}/\langle S/N \rangle$ (Caulet 1989), which
holds when the spectral resolution dominates over the line width, as is our
case. Since the spectra have increasing S/N with wavelength, there is no need
to define a maximum redshift for the sake of the rEW threshold. 

We found a total of $44$ systems with $0.015<W_0(2796)<2.028$ \AA, $4$ of them
with $W_0(2796)>1$ \AA\ (LOS 5, 6, 15, and 18). Out of these 4, two are
reported in the \mgii\ survey by PPB06 (see below), and two are new.

\subsubsection{\mgii\ in Low-resolution Spectra (Sample 'S2')}

Out of the 190 quasars in the SDSS-RCS sample, 144 form 375 pairs where a
\mgii\ system with $0.35<$\za$<0.9$ can be found. We shall call these quasars
the 'low-resolution sample', hereafter '{\it S2}'. Note that {\it S1} and {\it
S2} are {\it not} disjoint, since several quasars in {\it S2} were observed at
high resolution.

To find \mgii\ absorbers in {\it S2} we cross-correlated the sample with two
extant SDSS \mgii\ samples: the sample by PPB06, comprising 7421 absorbers
with $W_0(2796)>1.0$ \AA, and the sample by BMPCW06, made of 1806 absorbers
with $W_0>0.3$ \AA. PPB06 surveys the redshift range \za $\sim0.35-2.3$ and
BMPCW06 has \za $\sim0.37-0.8$.  Both samples resulted from searches in DR3
spectra. In the cross-correlation we imposed the criteria \za\ $\le
1.42$. This limit is given by the highest cluster redshift in the SDSS-RCS
sample (but note that we will later restrict the statistical samples to much
lower redshifts).

Out of the 144 quasars in {\it S2}, 22 are reported in PPB06 to show at least
one strong ($W_0(2796)>1$ \AA) \mgii\ system in the SDSS spectrum. Out of
these, one is in a quasar that is paired with a cluster at too low a redshift
and was therefore excluded. Out of the remaining 21 quasars in PPB06, two were
observed at high resolution and therefore are also included in {\it S1}.  The
remaining 19 quasars show 21 systems that are listed in~Table~\ref{tbl-qsos}
 along with absorption redshifts and rEW from PPB06
(LOS 20 and beyond). Let us emphasize that the two systems in LOS 15 and 18 of
{\it S1} are reported also by PPB06, so there is a total of 23 \mgii\ systems
with $W_0>1$ \AA\ in {\it S2} (in the LOS 15, 18, 20 and beyond) that were
reported by PPB06. The two other $W_0>1$ \AA\ systems in {\it S1} (LOS 5 and
6) are {\it not} reported in PPB06.

Out of the 144 quasars in {\it S2}, 5 are reported in BMPCW06 to show at least
one \mgii\ system with $W_0>0.3$ \AA\ in the SDSS spectrum.  Out of these, 4
with $W_0>1$ \AA\ are in the PPB06 sample (though 2 of these, 092746.94+375612
and 141635.78+525649, with rEW differing by $\approx 30$ \%) and only one has
$0.8<W_0<1.0$ \AA. We decided not to include this latter system into our
statistics because the redshift range surveyed by BMPCW06 is shorter than we
can probe with our quasar-redshift pairs. Therefore, only the PBB06 results
were used in our statistics. However, after calculating rEW values for the two
systems with disagreeging rEW in the two surveys, we decided ---for these two
particular systemsç--- to use the values reported by BMPCW06, which better
match ours (this choice has consequences for the rEW distribution below).

%

\subsection{Sample of Clusters and Survey Redshift Path}

\subsubsection{Cluster Redshift Intervals\label{intervals}}

Table~\ref{tbl-clusters} displays the cluster data for each LOS that contains
absorption systems (same numbering as in Table~\ref{tbl-qsos}). The 19 quasar
spectra in {\it S1} define a sample of 46 clusters with redshifts between
\zc$=0.173$ and $1.085$.  Out of these, 37 are drawn from the SDSS-RCS sample,
2 from the x-ray sample and 7 from the literature sample. In {\it S2} all
clusters come from the RCS.

RCS cluster redshifts are photometric and estimated to within $\delta z = 0.1$
in this redshift range (Gilbank et al. 2007)\footnote{For simplicity we have
firstly neglected the fact that the redshift accuracy of the RCS clusters is a
function of redshift, but address this later in \S~\ref{sect-redshift}.}. The
other 9 clusters have 
spectroscopic redshifts and we will assume $\delta z = 0.01$, which
corresponds to $\Delta v=2\,000$ \kms\ at \zc$\sim 0.5$. Since we will analyze
absorption systems with \za$\sim$\zc, our survey's redshift path will be
defined by what we shall call 'redshift intervals' around each quasar-cluster
{\it pair}. These are in turn defined as $[z_{\rm min},z_{\rm max}]$, with
$z_{\rm min}=$\zc$- \delta z$ and $z_{\rm max}=$\zc$+ \delta z$, unless
$z_{\rm min}<z_{\rm EW}$, in which case we set $z_{\rm min}=z_{\rm EW}$. This
choice implies that every redshift interval in {\it S1} permits a $>3\sigma$
detection of a system with $W_0>0.05$ \AA\ (this choice has also consequences
on survey completeness as explained below in \S~\ref{section-complete}). No
cluster has $z_{\rm max}<z_{\rm EW}$, so no redshift interval was excluded
from {\it S1}.  Recall that, in general, the number of redshift intervals is
{\it not} equal to the number of clusters, since some clusters are crossed by
more than one LOS.

For redshift intervals associated with {\it S2} we set $z_{\rm EW}=0.35$,
which defines a rEW threshold of $W_0^{\rm min}=1.0$ \AA. With this cut, out
of the 442 quasar-cluster pairs in the SDSS-RCS sample, 375 remain in {\it
S2}. These pairs are associated with 144 LOS. In Table~\ref{tbl-clusters} (LOS
20 and beyond) we show only clusters associated with the 19 quasars in {\it
S2}, besides LOS 15 and 18, that show a \mgii\ system with $W_0>1$ \AA.

\begin{figure}
\epsscale{.80}
\includegraphics[angle=0,scale=.8]{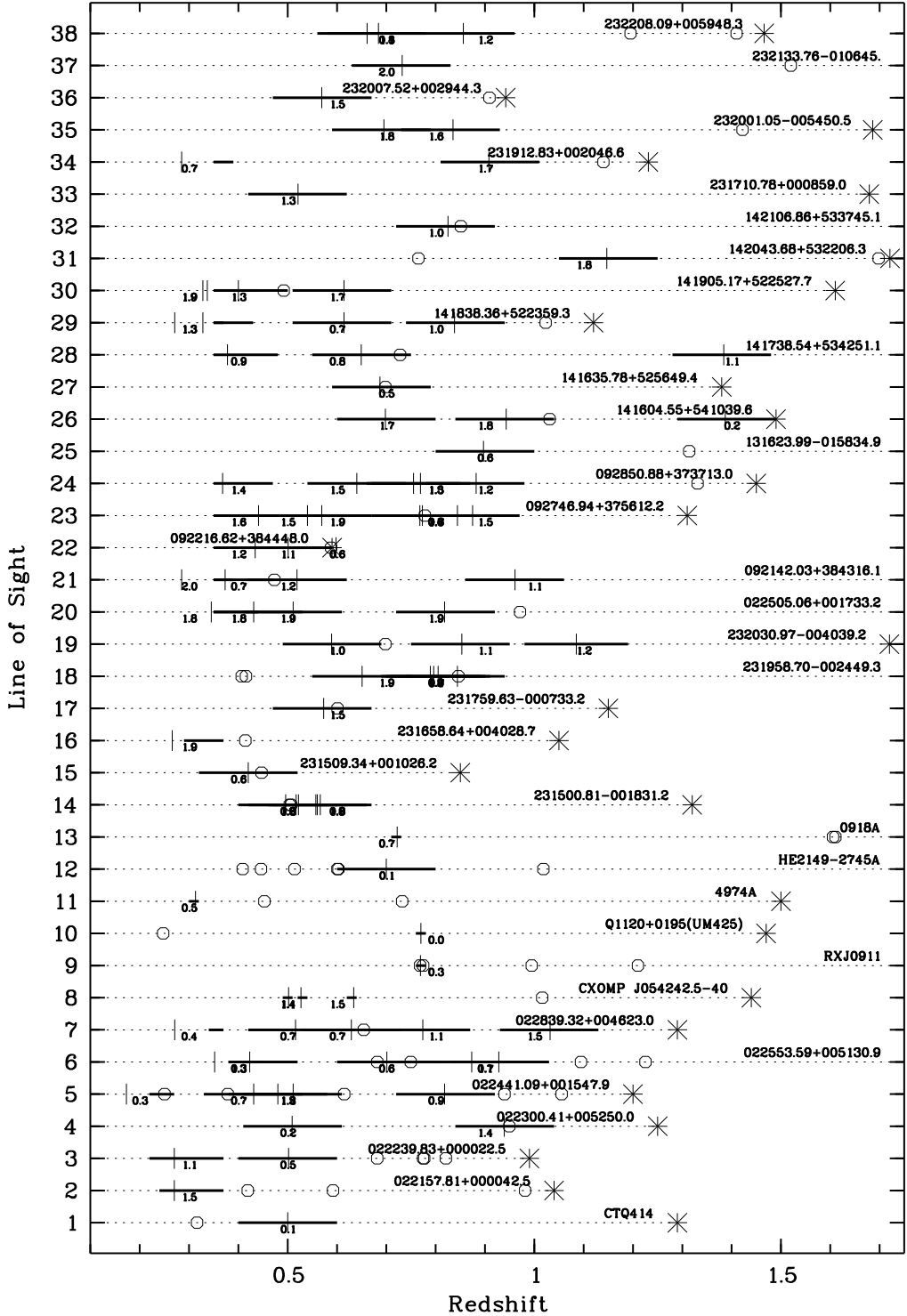}
\caption{ Diagram of the subset of lines of sight (LOS) toward which \mgii\
  absorption systems were found.  The LOS numbering is the same one used in
  Tables~\ref{tbl-qsos} and~\ref{tbl-clusters}. LOS up to 19 belong to sample
  {\it S1}; LOS 20 to 38 to sample {\it S2}. Quasar emission redshifts are
  labeled with asterisks, \mgii\ absorption systems with circles, and clusters
  with vertical lines. The thick lines despict the redshift intervals [$z_{\rm
  min},z_{\rm max}$] around cluster redshifts.  These intervals permit a $3
  \sigma$ detection of \mgii\ $\lambda 2796$ lines with $W_0>W_0^{\min}=0.05$
  \AA\ in {\it S1} and with $W_0>W_0^{\min}=1.0$ \AA\ in {\it S2}. The numbers
  below the thick lines are the projected LOS-cluster distance in 
  \hmpc\ at cluster redshift.
\label{fig_clusters}}
\end{figure}

Fig.~\ref{fig_clusters} shows a diagram of redshift intervals in each of the
LOS.  The LOS numbering is the same used in Tables~\ref{tbl-qsos}
and~\ref{tbl-clusters}.  Quasar emission redshifts are labeled with asterisks,
\mgii\ absorption systems with circles, and clusters with vertical lines. The
thick lines depict the cluster redshift intervals. The numbers below the thick
lines are the LOS-cluster distance (at $z_{\rm cluster}$) in \hmpc. LOS up to 19
belong to sample {\it S1}; LOS 20 to 38 to sample {\it S2}.

\subsubsection{A New Definition of Redshift Path Density\label{sect-g}}

In order to calculate the incidence of \mgii\ absorbers at cluster redshifts,
\zc, a function must be defined that accounts for the probability of detecting
the doublet at a given redshift.  In QAL surveys such a function is the
Redshift Path Density $g(W_{\rm min},z_i)$, which gives the number of
sightlines (quasar spectra) in which an absorption system with rEW
$W_0>W_0^{\rm min}$ might have been detected at redshift $z=z_i$ (see, for
instance, Eq. [1] in Ellison et al. 2004a).  Thus, in QAL surveys, $g(W_{\rm
  min},z_i)$ provides the redshift path sensitivity of the survey and the
total redshift path surveyed is given by:

\begin{equation}
\Delta z = \int_0^\infty g(W_{\rm min},z_i) dz.
\end{equation}

Since in the present analysis we are interested in the incidence of absorbers
at cluster redshifts, the following conceptual modification has to be
introduced: the redshift intervals defined in \S~\ref{intervals}, $[z_{\rm
min},z_{\rm max}]$, will be treated as if they were `quasar spectra', {\it
regardless} of how many of them are present in one LOS. The reason for this
choice is that having more than one cluster in the same LOS and at similar
redshifts (overlapping clusters) increases the {\it a priori} probability of
detecting an absorber in that particular LOS. Similarly, two different LOS
through the same cluster add twice to the overall redshift path.

We therefore define a `cluster redshift path density', $g_c(W_0^{\rm min},z_i,
  d)$, as the function that gives the number of cluster redshift intervals
  within a LOS-cluster distance $d$, in which a $W_0>W_0^{\rm min}$ \mgii\
  system at redshift $z_i$ might have been detected\footnote{Clearly, $g_c$ is
  also a function of $\delta z$, see \S~\ref{sect-redshift}.}. The cluster
  redshift-path $\Delta$\zc\ between any two redshifts $z_1$ and $z_2$ is thus

\begin{equation}
\Delta z_{\rm cluster}(W_0^{\rm min},z_1,z_2,d) = \int_{z_1}^{z_2} g_c(W_0^{\rm
  min},z,d)dz. 
\end{equation}

In Fig.~\ref{fig_g} we show $g_c(z)$ for the two samples.  Note that
$g_c(z)$ is not only different for each of the samples (because of
different rEW thresholds) but also for each cut in distance. Sample
{\it S1} provides a cluster redshift path between $z=0.2$ and $z=0.9$
of $\Delta$\zc$=6.3$ for $d<2$ \hmpc.  This is the longest path
available for searches of lines as weak as $W_0=0.05$ \AA. In the
redshift interval [0.35,0.90] and for $W_0>1$ \AA, sample {\it S2}
provides a redshift path of $\Delta$\zc$=57.0$ for $d<2$ \hmpc.  
 Overlaps represent $\approx 40$ \% of the total redshift path
for clusters at $d<2$ but only $\approx 10$\% for $d<1$ \hmpc. These
numbers are summarized in Table~\ref{tbl-samples}.

\subsection{Sample of \mgii\ absorbers at cluster redshifts: 'hits'}
  \label{stats}

We shall call an absorber a `hit' when \za\ is in a cluster redshift
interval. The function \nh=\nh$(z_1,z_2,W_0,d)$ is defined as the number of hits
between redshifts $z_1$ and $z_2$ with a given cut in rEW and distance.  \nh\
enters in the definition of $dN/dz$ below. We recall that (1) there may be
more than one hit in one redshift interval (two absorbers in the same LOS
through the same cluster); (2) there may be more than one hit in one cluster
(two absorbers in different LOS through the same cluster); and (3) redshift
intervals may overlap (thus increasing the probability of getting a hit).
Table~\ref{tbl-hits} summarizes the hits for the two samples and various cuts
in rEW and $d$.

The following caveat must be considered: overlapping redshift intervals have
no one-to-one correspondence with hits; in other words, we lack information as
of which one of the overlapping clusters is responsible for the
absorption.  This degeneracy, however, has a minor effect on the results
  by cluster impact parameter, since there are only two cases in the whole
  sample (LOS 5 and LOS 14) where a hit occurs in two overlapping intervals,
  with one being at $d<1$ and the other one being at $1<d<2$ \hmpc. These
  particular hits were assigned to both statistics: $d<1$, and $d<2$ \hmpc.

\subsection{Redshift Number Density of Absorbers in Galaxy Clusters}

To study the incidence of \mgii\ in cluster galaxies we define --- similarly
to an unbiased QAL survey defined by $W_0^{\rm min}$ --- the redshift number
density of absorbers in galaxy clusters, $(dN/dz)_c$, as the number of hits
per unit cluster redshift:

\begin{equation}
(dN/dz)_c (W_0,z_1,z_2) \equiv \frac{N_{\rm hits}(W_0,z_1,z_2)}{\Delta
  z_c(W_0,z_1,z_2)} , 
\end{equation}

and its rEW distribution, $n_c(W_0)\equiv \frac{d^2N}{dWdz}$, as the number of hits
per unit cluster redshift per unit EW, such that:

\begin{equation}
\int_{W1}^{W2} n_c (W_0,z_1,z_2)dW = (dN/dz)_c .
\end{equation}


The errors are calculated assuming Poisson statistics, for which we use the
tables in Gehrels (1986).

These two observational quantities, $(dN/dz)_c$ and $n_c(W_0)$ must be
proportional 
to the average number density of absorbers in a cluster, $n_c(z)$, and their
cross-section, $\sigma_c(z)$:

\begin{equation}
(dN/dz)_c  \propto n_c(z)~\sigma_c(z)~.
\label{eq-cosmo}
\end{equation}


Although in general $(dN/dz)$ has been used to study how absorbers evolve,
our samples are rather small and we just focus on a possible overdensity
$\delta$ of
absorbers with respect to the field. We define
\begin{equation}
\delta\equiv (dN/dz)_c/(dN/dz)_f~ ,
\end{equation}
where $(dN/dz)_f$ is the incidence of systems in the field. 
We compare the two distributions
measured in clusters with the following field \mgii\ surveys: NTR06 (MMT
telescope spectroscopy, spectral resolution FWHM $\approx 2.2$ \AA; rEW
threshold $W_0^{min}=0.1$ \AA), NTR05 (SDSS EDR, FWHM $\approx 4$ \AA,
$W_0^{min}=0.3$ \AA), CRCV99 (Keck HIRES, FWHM $\approx 0.15$ \AA,
$W_0^{min}=0.02$ \AA), and Narayanan et al. (2007; VLT UVES, FWHM $\approx
0.15$ \AA, $W_0^{min}=0.02$ \AA).  Other surveys have redshift intervals that
do not match ours (Lynch, Charlton
\& Kim 2006).

These surveys have found (1) that weak and strong systems show different rEW
redshift distributions: weaker systems are fitted by a power-law while
stronger systems are better described by an exponential, with the transition
at $W_0\approx 0.3$ \AA. This effect would hint at two distinct populations of
absorbers (e.g., NTR05); (2) little evolution of any of the populations
between $z\approx1.4$ and $0.4$ (Narayanan et al. 2007; Lynch, Charlton \& Kim
2006). The nature of weak ($W_0<0.3$ \AA) \mgii\ is not clear yet.  It has
  been suggested that single-cloud systems may have an origin in dwarf
  galaxies due to their abundances (Rigby et al. 2002) or to their statistics
  (Lynch, Charlton \& Kim 2006); they might also be the high-redshift analogs
  to local HVCs (Narayanan et al. 2007, and references therein).
  Unfortunately, there exist only few QAL surveys of weak \mgii\ systems,
  mainly due to the more scarce high-resolution data.

\section{Results: The Incidence of \mgii\ in Galaxy
  Clusters\label{sect-results}}  

In this section we present the results on $(dN/dz)_c$ and $n_c(W_0)$ as
observed in {\it S1} (for systems having $W_0<1.0$ \AA) and {\it S2}
($W_0>1.0$ \AA). For both samples we analyze pairs with $d<2$ and $<1$
\hmpc\ separately, and we restrict the statistics to $z<0.9$, where the
cluster sample is more reliable. Finally, we re-analize {\it S2} taking into
account two refinements of the method, namely selection by cluster richness
and the redshift-dependence of $\delta z$.

\subsection{$W_0<0.3$ \AA\  systems}

The parameterization by CRCV99 of their Keck HIRES data implies
$(dN/dz)_f=1.41$ at $\langle z\rangle=0.65$ for field systems with
$0.02<W_0<0.3$ \AA\ at $0.4<z<1.4$. This is consistent with the results by
Narayanan et al. (2007) at that redshift and in the same rEW interval using
UVES data.

For  our redshift intervals having $d<1$ \hmpc\ we find $(dN/dz)_c=1.20$ ([0.37
  2.70] $1\sigma$ c.l.) for $0.05<W_0<0.3$ \AA\ and binning in the range
$0.2<z<0.9$.  Given that our data are complete only down to $W_0=0.05$ \AA, we
cannot compare directly with the value by CRCV99. Therefore, we apply a
downward correction to this value of $23.3$ \%, which is the fraction of
systems with $0.02<W_0<0.05$ \AA\ in the CRCV99 sample. After this correction,
the field value is $(dN/dz)_f=1.09$, which is in good agreement with
$(dN/dz)_c$.  For the $d<2$ \hmpc\ sample we find a somewhat smaller value of
$(dN/dz)_c=0.79$ ([0.29 1.64] $1\sigma$ c.l.)  that is however still
consistent with the field measurement.

\subsection{$W_0>0.3$ \AA\  systems}

Figure~\ref{fig_n_z} shows the cumulative values of $(dN/dz)_c$ (and their
$1\sigma$ errors) for systems with $W_0>0.3$ \AA. We bin in the ranges
$0.2<z<0.9$ (top panels) and $0.35<z<0.9$ (bottom panels).  The top panels
show results from sample {\it S1} only (46 quasar-cluster pairs,
\zave$=0.550$), while points in the bottom panels were calculated using sample
{\it S2} (375 pairs; \zave$=0.625$). The filled circles are for clusters at
distances $d<2$ \hmpc\ from quasar LOS and the open squares represent clusters
with $d<1$ \hmpc\ (symbols are slightly shifted in the x-axis for more
clarity).  The curves correspond to the fit by NTR05 to their EDR data of
field absorbers with $1\sigma$ limits calculated as described in the Appendix
of their paper. These fits are in excellent agreement with the SDSS data of
field \mgii\ absorbers.

There is an overdensity of hits per unit redshift in clusters compared with
the field population for $d<1$ \hmpc\ clusters; the $d<2$ \hmpc\ sub-samples
instead, are consistent with the field statistics. In addition, the data show
that $\delta$ is larger for stronger systems ($W_0>2.0$ \AA) than for weaker
systems.  These two trends are more clearly seen in Table~\ref{tbl-hits},
where we compare the measured value of $(dN/dz)_c$ with the field, for various
rEW ranges (using cosmic averages from different authors). Note that the
confidence limits listed in the Table for $(dN/dz)_c$ are $2\sigma$ only. The
overdensity effect for $d<1$ \hmpc\ ($W_0>1$ and $W>2$ \AA\ cuts) is
significant at the 99\% level or slightly higher. For $d<0.5$ \hmpc\ we also
note the overdensity of stronger systems, though the effect here is only
$1\sigma$ due to the small number of hits.


\subsection{\mgii\ $\lambda 2796$ equivalent width distribution}

\subsubsection{Stronger (Weaker) Systems in Clusters are (not) Overdense} 

Fig.~\ref{fig_n_W} summarizes our main result. It shows $n_c(W_0)$ and
$1\sigma$ errors for \mgii\ systems at $d<1$ and $d<2$ \hmpc\ from a cluster.
Data points with $W_0<1.0$ \AA\ result from sample {\it S1} only, while points
at $W>1$ \AA\ are calculated using {\it S2} only. The solid curve is the
exponential distribution $n(W_0)=N^*/W^*\exp{-W_0/W^*}$ fitted by NTR06 to
their MMT data for $W_0>0.3$ \AA\ (114 \mgii\ systems, \zave$=0.589$). The
parameters are $W^*=0.511$ and $N^*=1.071$ and the fit is in excellent
agreement with their data having $0.5\la W_0\la 3.0 $ \AA\ (see their Fig. 2).
The dashed curve is the power-law fit, $n(W_0)=0.55~W_0^{-1.04}$ fitted by
CRCV99 to their Keck HIRES data. The power-law is in excellent agreement
with their $W_0\la 0.3$ \AA\ data and also with data by Steidel \& Sargent
(1992), but clearly overestimates the MMT and SDSS/ERD data for larger $W_0$.

Fig.~\ref{fig_n_W} confirms the excess of {\it strong} ($W_0\ga 1.0$ \AA)
\mgii\ systems near cluster redshifts, when compared with the field
population.  On the contrary, the weaker systems ($W_0\la 0.3$ \AA) conform to
the field statistics. Furthermore, this effect seems more conspicuous for the
$d<1$ \hmpc\ sample than for the $d<2$ \hmpc\ sample, which shows a slight
overdensity only for stronger systems.  The weak systems are also consistent
with other QAL surveys. For instance, the results by Narayan et al. (2006) for
the $0.4<z<1.4$ range are in good agreement with ours (their Fig.~7) even for
our $d<1$ \hmpc\ sample, considering a $29.3$ \%\ downward correction to their
$n(W_0)$ values due to our smaller rEW range of [0.05,0.3] \AA.  On the
contrary, for $d<1$ \hmpc, $n_c(W_0)$ is overabundant by a factor of $\approx
3$ in the [1.0,2.0] bin and $\approx 15$ in the [2.0,3.0] bin (note
Table~\ref{tbl-hits} shows a comparison with NTR05). The latter result is
significant at the $> 3\sigma$ level (assuming no errors in the field
average).

Summarizing, stronger systems ($W_0\ga1.0$ \AA) are overdense in clusters;
weaker systems ($W_0\la0.3$ \AA) are not.  This makes $n_c(W_0)$ appear flatter
than $n(W_0)$ (on a log-log plot) and much better fitted by a power-law, {\it
also in the large-rEW end}, than by an exponential.

\subsubsection{Is the Effect Real?}

The different behaviour of strong and weak systems cannot be due to the
different redshift paths $\Delta z_c$ of {\it S1} and {\it S2}.  If this were
the case, the offset in $(dN/dz)_c$ should be equal in the entire rEW range;
however, we see that the statistics is affected differentially.

Another possible caveat is that an incomplete survey in the small-rEW end
would as well have an effect on the differential behaviour of $n(W_0)$ between
weak and strong systems (weaker lines are more difficult to find). However,
the [0.05,0.3] bin has 4 absorbers in {\it S1}, meaning that to get a factor
of say $10$ more systems per unit rEW in that bin we should have missed 36
hits, which is quite unlikely ($10^{-12}$ for a Poisson distribution).

The stronger effect seen for $d<1$ when compared with $d<2$ \hmpc\ is clearly
influenced by the shorter redshift path in the former selection. Indeed, from
Table~\ref{tbl-hits} we see that the transition from $d<2$ to $d<1$ \hmpc, is
more or less governed by the change in $\Delta$\zc\ (i.e., the number of hits
do not change much). We take this as a possible evidence that the data is
sensitive to a typical cluster only at distances below $1$ \hmpc. Further
support for this idea is that the $d<0.5$ \hmpc\ overdensities, though at low
significance, do not scale with $\Delta$\zc. 

 Finally, let us note that our definition of $\delta$ and the large radial
  distances implied by the photometric redshift accuracy ($\delta z$) imply
  that $\Delta z_{\rm cluster}$ may (and probably does) include some level of
  contamination by field absorbers. Consequently, what the present cluster
  data allows us to state is that {\it regions} that contain a cluster do show
  more strong absorbers than the cosmic average, while for weak systems those
  regions are indistinguishable from the field.

%

\subsection{Refinements}

\subsubsection{Selecting by Cluster Richness\label{section-rich}}

In the current analysis we have included all the objects in our RCS catalog,
constraining its significance to be greater than $3\sigma$ (Gladders and Yee
2000). This threshold is low enough to detect almost all the clusters in the
RCS areas, but presumably it also includes low mass groups and even some
spurious detections. On the other hand, this selection has the great feature
that allows us to cross correlate a large number of objects, but also it has
the drawback that any signal we detect in the absorption systems might be
diluted by low mass objects or spurious clusters.

In order to quantify the extent of this 'dilution' we have selected a
subsample of clusters with a more stringent criteria given by a minimum
richness (that translates into a minimum mass). So far we have used sample {\it
S2} that has a median $B_{gc}$ of 263 that translates into a mass of
$\approx2.4 \times 10^{13} M_{\odot}$ (Blindert et al., in prep.). Similar values
are obtained for the subsample having $d<1$ \hmpc. 

The more restricted sample, which we call '{\it S2-best}' is required
to have only clusters with $B_{gc} \ge 350$. This selection includes
only 125 quasar-cluster pairs for an impact parameter of at least 2
\hmpc, and a median $B_{gc}$ of 488 that translates into a mass of
$8.8 \times 10^{13} ~M_\odot$. Similarly, we find a median $B_{gc}$ of
478 for a $d<1$ \hmpc.  As shown in Table~\ref{tbl-hits}, repeating
the analysis of hits using {\it S2-best} yields higher overdensities
---by $\sim 50$\%--- than for {\it S2} for the same rEW ranges. Most
importantly, despite a much lower redshift path, the {\it
significance} of the result is still high ($>3\sigma$).

Sample {\it S1} has fewer quasar-cluster pairs and only a few of the
clusters come from the RCS sample. For these objects we find a median
$B_{gc}$ of 327 for an impact parameter of 2 Mpc and 258 for the
smaller aperture, i.e., consistent with the larger sample.  Therefore,
a similar analysis for {\it S1} was considered not worth performing
due to the few clusters in that sample.  However, we note that the
median $B_{gc}$ is not particularly low, so this sample is
also representative of more massive clusters (i.e., the lack of
overdensity is not due to a chance conjunction of low mass systems).

Concluding, finding a significantly stronger signal in clusters
selected by a mass proxy gives strong support to both the method and
the reliability of the systems used in the analysis. In fact, the
richness selection not only provides more galaxies per clusters but
also picks up larger clusters. Both selection effects are expected to
increase the a priori hit probability.

\subsubsection{Cluster Redshift Accuracy\label{sect-redshift}}

 Since our comparison between cluster and field \mgii\ statistics depends
  on the definition of redshift intervals, we have to consider what effect the
  cluster redshift accuracy may have on our results. For clusters with
  spectroscopic redshifts we have assumed $\delta z=0.01$. If due to the
  Hubble flow, this translates into a radial distance of $67$ comoving Mpc;
  therefore, one might want to shorten $\delta z$ to overcome the problem of
  contamination by field absorbers. Unfortunately, the redshift path provided
  by the pairs with spectroscopic cluster redshifts represents only $\approx
  2$ \%\ of $\Delta z_{\rm cluster}$.  
Since
  shortening to $\delta z=0.005$ does not exclude the only hit (LOS 9) at a
  spectroscopic $z_{\rm cluster}$, $(dN/dz)_c$ remains practically unchanged.

The vast majority of our clusters have photometric redshifts and  our
analysis assumes $\delta z=0.1$ for those ones.  This is indeed an
over-estimate for the lower-redshift clusters, \zc$\la 0.5$, where the
accuracy can be as good as $\delta z\approx0.04$. In order to see whether a
smaller $\delta z$ would affect the results on $(dN/dz)_c$ we use the
parameterization $\delta z = 0.04 (1+z_{\rm cluster})$ and re-compute $g_c(z)$
and $N_{\rm hits}$. Restricting the analysis to $d<1$ \hmpc\ pairs in {\it S2}
(where the overdensity signal is most evident), we find that out of $7$ hits
with $W_0>1$ \AA\ only one hit (LOS 28, $W_0=1.58$ \AA) is ruled out due to
the shorter redshift intervals. Since the new $\delta z $ makes the total
redshift path between $z=0.35$ and $z=0.9$ decrease to $\Delta z_{\rm
  cluster}=9.18$, we find actually a higher overdensity of $\delta\approx 4$
and $\delta\approx 10$ for $W_0>1.0$ and $1.0<W_0<2.0$ \AA, respectively. We
conclude that our result is indeed affected by a more precise parameterization
of the RCS redshifts but such refinement makes the signal even stronger. In
order to avoid fine-tuning too many variables, we continue the analysis of the
results using a constant $\delta z$.

\section{Statistical Significance, Possible Biases, and
  Caveats\label{sect-stats}} 

Despite the strong test provided by the mass selection, our method might still
suffer from possible systematics and biases hidden in the statistical
properties of the various samples. We analyze these in what follows.

\subsection{Statistical Significance}

We start by asking whether the detected overdensity might be due to chance
alignments.  To assess the statistical significance of the observed number of
hits one might want to run Monte Carlo simulations by creating samples of
random cluster redshifts.  However, this is equivalent to calculating $dN/dz$
from random sub-samples drawn from the parent quasar sample (i.e., creating
random RCS-SDSS samples). Such kind of simulations must, by definition, yield
the cosmic value obtained by QAL surveys, a number against which we have
compared our resuls.
%
To see whether we recover the expected number of field absorbers, we calculate
$(dN/dz)_f$ in the {\it complementary} redshift path of our quasar-cluster
sample, that is, the path that does not include clusters. If our sample is
biased toward an overdensity of absorbers for reasons {\it other} than the
presence of clusters we should get an overdensity here too; if it is not, we
should recover the field value. We analyze quasars in sample {\it S2}, the one
that yields the overdensity, and split it into two redshift ranges:
$z=$[0.35,0.9], the one used to get $(dN/dz)_c$, and $z=$[0.9,1.4]. The latter
was not used in the analysis of cluster absorbers but may be a useful check
for unbiased LOS. 

There are $85$ quasars in {\it S2} that provide cluster redshift intervals at
$d<1$ \hmpc, where $W_0>1$ \AA\ \mgii\ systems may be detected. Between
$z=0.35$ and $0.9$ the total {\it quasar} redshift path of this sample is
$\Delta z_{\rm quasar} =45.15$, so the complementary redshift path is $\Delta
z_{\rm field} = \Delta z_{\rm quasar} - \Delta z_{\rm cluster} = 45.15 -
14.13 = 31.02$, where we have subtracted the cluster path $\Delta$\zc$=14.13$
(see Table~\ref{tbl-hits}).

The expected number of $W_0>1.0$ \AA\ systems along $\Delta z_{\rm field}$ is
thus $5_{-2.2}^{+3.4}$ and the expected number of systems with $2.0<W_0<3.0$
\AA\ is $0_{-0}^{+1.9}$ ($1\sigma$ errors).  From Tables~\ref{tbl-qsos}
and~\ref{tbl-hits}, the observed number of systems along $\Delta z_{\rm
field}$ in this redshift range is: [\# systems in RCS-SDSS] $-$ $N_{\rm hits}$
= $10-7=3$ for $W_0>1.0$ \AA\ and $3-3=0$ for $2.0<W_0<3.0$ \AA\ systems
(i.e., no system with $2.0<W_0<3.0$ \AA\ was expected in the field and no
system was observed in the field, with the 3 other systems all being
hits). These values are in agreement with the field expectation.

Repeating the above analysis for the $z=$[0.9,1.4] range, we get: $\Delta
z_{\rm quasar} =31.38$, $\Delta$\zc$=3.00$, number of expected field
absorbers: $7_{-2.6}^{+3.8}$, number of detected field absorbers: 8 (total) -
1 (hit) = 7, i.e., again within the field expectation.  We conclude that the
observed overabundance of strong \mgii\ systems is not due to chance
alignments and must reflect real overdensities.  In other words, sample {\it
S2} of {\it quasars} is biased {\it only} by the presence of clusters. The
bias vanishes at redshifts other than \zc, where we recover the cosmic
statistics obtained in QAL surveys (the presence of clusters in these surveys
has negligible influence on such statistics).

\subsection{Yet More Possible Biases and Caveats\label{section-complete}}

\subsubsection{Clusters and Quasars}

Besides the obvious fact that the completeness of our cluster sample
--- drawn mainly from the RCS --- depends on the RCS algorithm, it is
important to stress that the parent cluster and quasar samples are
totally independent each from the other. The RCS sample is certainly
not complete for {\it S2} (particularly at higher redshifts), which
includes low mass systems, but it is for {\it S2-best} up to $z\la 1$,
which includes moderately massive clusters. On the other hand, the SDSS
quasar sample should be $\sim 90$ \% complete (York et
al. 2000). These and the arguments given in \S~\ref{sect-correlation}
lead us to conclude that the SDSS-RCS sample, and thus also sample
{\it S2-best} of pairs, is complete and homogeneous, at least at the
same level as their parent surveys.

Another obvious strength of the quasar-cluster sample is that the search of
absorbers in {\it S2} (PBB06; BMPCW06) was  performed independently of our
selection. This is not completely true for {\it S1} since those quasars were
selected for follow-up spectroscopy {\it after} the quasar-cluster
selection. However, at the telescope, the targets were selected without prior
knowledge of cluster redshifts; moreover, even if this had been the case, the
low-resolution spectra provided by the SDSS do not permit an a priori
selection of {\it weak} systems.  Therefore, there was no way to prefer
quasars with absorbers. We discuss this further below in the context of
absorber statistics.

Finally, the following caveat must be mentioned: brighter quasars are
chosen for spectroscopy, which might be amplified by gravitational lensing by
the absorber host galaxies (see discussion in \S~\ref{sect-discussion}).

\subsubsection{Absorbers}

Surveys of quasar absorption-line systems assess the completeness of the
samples via cumulative $\Delta z$ as a function of rEW threshold (Steidel et
al. 1992). Since we have {\it chosen} our redshift path to include only
spectral regions sensitive to $W_0>0.05$ \AA, we consider the sample {\it S1}
of absorbers to be nearly 100\% complete. Similarly, we assume that {\it S2}
is statistical in the sense that all \mgii\ systems with $W_0>1.0$ \AA\ were
listed in PPB06, who argue that their search is $>95$ \% complete.

As for the homogeneity of the samples, we have kept {\it S1} and {\it S2}
carefully separated. Again, out of the 4 $W_0>1$ \AA\ systems found in {\it
S1}, we have considered in {\it S2} only those two found by PBB06 (including
the remaining two would increase $\delta$ since one system is a hit).

Admittedly, one concern is that detecting an overdensity in one sample and not
in the other may reflect a hidden systematic. We do not have at this time the
means of testing such possible systematics. If we use only {\it S1} in the
$W_0>1$ \AA\ range we also find an overdensity with respect to the field,
although with low significance: $1$ hit is expected while $2$ are found.
However, as pointed out above, these statistics may be influenced by the fact
that quasars in {\it S1} were selected as having a cluster in the LOS, and
strong systems are readily seen in the SDSS spectra. However, weak systems are
not seen in the SDSS and we know they do not cluster around stronger systems
(CRCV99). In addition, if there were in fact such hidden systematics, why is
not the supposedly biased sample ({\it S1}) the one that shows the
overdensity? In other words, despite an obvious selection effect
toward targets with clusters, {\it S1} {\it does} yield the field statistics
for weak absorbers.

 Finally, note that the significance of our result for strong absorbers could
increase if a larger cluster redshift path were surveyed.
Tables~\ref{tbl-samples} and ~\ref{tbl-hits} show that only roughly $3-4$\% of
the quasar-cluster pairs results in hits. This explains why an earlier attempt
failed to detect strong \AA\ \mgii\ systems in a sample of 6 Abell
clusters (Miller, Bregman \& Knezek 2002).

\section{Summary of the Results\label{sect-summary}}

We have cross-correlated candidate galaxy clusters from the RCS at
\zc$=0.3$--$0.9$ with background quasars from the SDSS DR3 to investigate the
incidence $(dN/dz)_c$ and rEW distribution $n_c(W_0)$ of \mgii\ absorption
systems associated with cluster galaxies. We have found 442 quasar-cluster
pairs at impact parameters $d<2$ \hmpc\ from cluster coordinates, where
\mgii\ might be detected  in redshift regions $\pm 0.1$ from a
  cluster. The cluster sample contains all systems in the RCS, and is
dominated by low-mass clusters and groups with $\langle M \rangle \sim
2\cdot10^{13}$ M$_\sun$ cluster candidates. We have defined a cluster
redshift-path density in terms of the quasar-cluster pairs. Using extant
surveys of strong \mgii\ systems in DR3 quasar spectra and our own follow-up
high-resolution spectroscopy, we calculated $(dN/dz)_c$ and $n_c(W_0)$ for the
rEW range $0.05<W_0<3.00$ \AA. The results were:

\begin{enumerate}

\item 
There is an excess of {\it strong} ($W_0^{2796}>1.0$ \AA) \mgii\
absorbers near ---i.e., at similar redshift of and close in projection
to--- galaxy clusters when compared to surveys in the field. The
effect is significant at the $2\sigma$ level. This overdensity,
$\delta\equiv (dN/dz)_c/(dN/dz)_f$, is also more pronounced at smaller
distances ($d<1$ \hmpc) than at larger distances ($d<2$ \hmpc) from
the cluster, which we interprete as a dilution of the effect in the
field.  On the other hand, the excess is also more pronounced for
stronger systems.  For $d<1$ \hmpc\ and $W_0=$ [2.0,3.0] \AA, we
measure $\delta\approx 6$--$15$ (depending on the field survey used
for comparison), and the significance increases to $3\sigma$.

\item 
If we select the sample third with most massive (and significant) cluster
candidates, we find the excess of absorbers increases by 50\%
for the sub-sample dominated by $\langle M \rangle \sim 10^{14}$ M$_\sun$
clusters. The effect becomes also more significant, rendering reliability to
our detection.

\item
The {\it weak} population of \mgii\ systems ($W_0<0.3$ \AA) in clusters
conform to the field statistics. The absence of an overdensity is not due to
lack of sensitivity. This effect and the excess of strong systems make
$n_c(W_0)$ appear flatter on a log-log scale, so ---contrary to the field---
it can be fitted by a power-law over the whole range of rEW.

\end{enumerate}

\section{Discussion\label{sect-discussion}}

The most obvious interpretation for the observed overdensity of strong
absorbers is that clusters represent a much denser galaxy environment
than the field: an overdensity is expected if field and cluster
galaxies share the same properties responsible for the \mgii\
absorption.  Below we discuss this possibility and then hypothesize on
why this trend is seen only for the strong cluster absorbers (thus
producing a flatter rEW distribution than in the field). Finally, we
assess the implications for the fraction of cold gas in galaxy
clusters.

One evident caveat to have in mind in comparing cluster and field properties
of \mgii\ is that possible correlations between rEW and galaxy properties
(colors, absorber halo mass, dust redenning and gravitational lensing) all
have been studied in the framework of the overall population of absorption
systems. Such field properties do not necessarily hold for clusters, and any
departure due to cluster environments may have not been detected in the field
studies.

\subsection{Galaxy Overdensity}

In order to see whether the observed overdensity of  strong absorbers,
$\delta$, is consistent with a model of evolution of structure a detailed
numerical simulation is necessary, which is out of the scope of the present
paper and will be presented elsewhere (Padilla et al., in prep.).  However, a
crude estimate of the expected $(dN/dz)_c$ can be obtained from
semi-analytical models. We start assuming field and cluster absorbers share
the same \mgii\ cross section, $\sigma$. In this case, from
Eq.~\ref{eq-cosmo}, $\delta$ is proportional to the average volume overdensity
of galaxies in a cluster, $\delta_g$.  To calculate $\delta_g$, we assume that
galaxies are spatially distributed in the same way as the dark matter, and
therefore adopt a NFW density profile (Navarro, Frenk \& White, 1997), which
depends on the total mass of the cluster of galaxies.  We then calculate
$\delta_g$ within $1$ and $2$~\hmpc\ from the cluster center, for masses
corresponding to the range present in the RCS sample.  Given that the LOS will
actually cross different density amplitudes as it passes through a cluster, we
simply make an order of magnitude approximation and take half the actual
overdensity at the impact parameter.  The number density of galaxies within a
cluster of galaxies is obtained assuming a Halo Occupation number
corresponding to a magnitude limit of $M_{r(AB)}=-17$, which states that the
number of galaxies populating dark-matter halos of a given mass is (Cooray
2006)
\begin{equation}
N=\frac{1}{\exp(\beta*(M_{min}-M))+1}+10^{\alpha*(M-M_1)},
\end{equation}
whith $M_{min}=10^{11.77}h^{-1}M_{\sun}$, $M_{1}=10^{12.96}h^{-1}M_{\sun}$,
$\alpha=1.04$, and $\beta=99$.  We then calculate the average density of
galaxies above the same magnitude limit by populating all haloes in the
Millenium simulation (Croton et al. 2006) using this same prescription, and
then counting the total number of galaxies and dividing this by the total
volume of the simulation.  

 The results for later-type galaxies are displayed in
  Table~\ref{tbl-galaxy}. Note that this estimate for $\delta_g$ includes only
  the cluster region; thus, it is to be compared with $\delta -1$, i.e., the
  overdensity of absorbers after subtracting the field contribution.

\subsection{$W_0\ga1.0$ \AA\ Absorbers}

\subsubsection{Absorber Overdensity}
\label{overdensity}

If we first concentrate on moderate mass clusters, $M\sim2\cdot10^{13}
M_\sun$, which vastly dominate sample {\it S2}, we see the expected
overdensity of cluster galaxies  is quite in line with $\delta-1$, the
  observed absorber enhancement, for $d<1$ and $d<2$ \hmpc\ (using for {\it
  S2} fiducial values of $\delta = 2$ and $\delta =10$ for the two apertures,
respectively; see Table~\ref{tbl-hits}).  In other words, the probability of
hitting a \mgii\ galaxy in a cluster is the same as in the field. Note that
this does not imply that quasar LOS do not 'see' the foreground clusters but
rather that this probability scales with galaxy overdensity.  The observed
match  between $\delta_g$ and $\delta -1$ supports the hypothesis that
strong absorbers in less masive clusters and in the field share similar
properties.

The situation seems different for sample {\it S2-best}, which is dominated by
more massive, $M\sim10^{14} M_\sun$, clusters. There we find that the absorber
overdensity is enhanced by a factor of $\la 2$ with respect to that one in
less massive clusters.  On the other hand, from Table~\ref{tbl-galaxy} we see
that the more massive clusters provide a factor of 5 more galaxies than less
massive ones. Therefore, the overdensity of galaxies in massive clusters
overpredicts the overdensity of strong absorbers by a factor of $\approx
2$-$3$.  We infer that ---on average and neglecting other effects, see next
Section--- the total cross-section of strong absorbers must be smaller in more
massive clusters than in the field by a factor of $\approx 2$-$3$.

It is tempting to draw conclusions also for $d<0.5$ \hmpc, despite the less
  significant signal observed in the absorber statistics. For both mass
  ranges, the expected overdensity of galaxies increases by a factor of
  $\approx 4$ comparing the $<1$ \hmpc\ and the $<0.5$ \hmpc\ apertures.
  Again assuming same properties as in the field, the absorbers statistics
  fails to reproduce such increase by that same factor of 4 (since
  roughly the same $\delta$ is observed for $d<1$ and $d<0.5$ \hmpc). This
  could indicate the gas cross section is even smaller at distances closer
  than half Mpc to the cluster center (although the factor $4$ is still within
  $95\%$ confidence limits).

%
%

\subsubsection{Gravitational Lensing}

Although we will present elsewhere a study of gravitational lensing by the
cluster galaxies in our sample, this effect
deserves a few words here since it may have direct implications for
$(dN/dz)_c$.  We are particularly interested in lensing magnification by the
RCS galaxies and the possible bias it may have introduced in the SDSS sample
used here.  Inclusion of magnified quasars in magnitude-limited surveys like
the SDSS might increase the number of (lens) absorbers per unit redshift
(1997 Bartelmann \& Loeb 1996; Smette, Claeskens \& Surdej).

Lensing magnification has been reported not to induce a significant
effect on the field statistics of strong \mgii\ systems (as observed
in SDSS quasar spectra; Menard et al. 2007). However, in our case the
probability of strong lensing might be greatly enhanced due to not
only the quasar light crossing the densest galaxy environments, but
also to a possible combination of cluster/quasar redshift ratios of
1:2 that maximizes the probability of strong lensing for $z_{\rm
em}\sim 1$ (indeed, that probability is maximal at $z_{lens} \sim 0.7$
for $z_{\rm em}\ga 2$).  Statistical overdensities of bright
background quasars (or paucity of faint quasars) associated with
foreground clusters or large structure have been already detected
(Myers et al. 2003; Scranton et al. 2005; but see Boyle, Fong \&
Shanks 1988). However, when searching for the lensing galaxies, one
finds that the majority of them are early-type (e.g., Fassnacht et
al. 2006) which are not expected to host strong \mgii\ absorbers
(Zibetti et al. 2007). We conclude that a great impact of lensing on
$\delta$ should not be expected. Nevertheless, if a fraction of these lensing
galaxies indeed does act as strong \mgii\ absorbers, then $(dN/dz)_c$
observed in our sample might be partly due to lensing. In such a case,
the values quoted in \S~\ref{overdensity} for the fractions of \mgii\
cross section that is expected from galaxy counts but not observed in
absorption must be seen as upper limits, since they result from a
sample that is biased toward more lensing absorbers.




\subsection{$W_0\la 0.3$ \AA\ Absorbers}

\subsubsection{A flatter rEW Distribution for clusters}


In contrast with strong systems, we do not detect an overdensity of weak
absorbers in clusters, although our survey is sensitive enough in the
$W_0<0.3$ \AA\ range.  As already stated, this dichotomy induces a flatter, more
uniform, rEW distribution than what is observed in the field, where weak
absorbers have a much steeper distribution than strong absorbers.  Does this
mean that neither gravitational lensing nor galaxy overdensity influence the
weak absorber statistics?  Since lensing magnification is a strong function of
\mgii\ rEW (Bartelmann \& Loeb 1996), $(dN/dz)_c$[$W_0<0.3$\AA] is perhaps
insensitive to lensing. However, it would be unlikely that also the galaxy
excess that clusters represent had no influence on the incidence of the weak
absorbers. This would require a physically distinct population of cluster
absorbers, detached from the strong absorbers, that does not scale with
galaxy overdensities.

Alternatively, since either the absorber number density or
filling-factor/cross-section affect the $(dN/dz)_c$ statistics, an interesting
possibility is that we have detected the signature of {\it processes} giving
rise to \mgii\ absorption (gas outflows or extended halos, the two current
compelling scenarios) that are at play in clusters in a different way than in
the field. For instance, if we consider the extended halo hypothesis
(Churchill et al. 2005), the low rate of weak \mgii\ absorbers we observe in
clusters might be due to truncated halos due to environmental effects.  Such
an effect is expected if cluster galaxies lose their gas after a few orbits in
processes like galaxy harassment and/or ram pressure stripping (Mayer et
al. 2006), and it has actually been observed in 21cm observations of
low-redshift (Giovanelli \& Haynes 1983; Chung et al. 2007; Verheijen et
al. 2007) and  local (Bravo-Alfaro et al. 2000) clusters.  Interestingly,
ram pressure affects mostly less-massive galaxies; on the other hand,
according to some authors weaker systems seem to arise in under-luminous,
less-massive galaxies (Churchill et al. 2005; Steidel et al. 1992). All this
fits 
well with the lack of absorbing cross-section observed here for cluster
galaxies associated with weak \mgii\ absorption.

If we instead consider the weak absorbers to be individual, small
'clouds' that are distributed more densely toward the centers of
galaxies, then weak \mgii\ arises in sightlines through the outer
parts of a galaxy (e.g. Ellison et al. 2004b). This is supported
by the typical sizes of strong/weak \mgii\ which are an order of
magnitude different (Ellison et al. 2004b).
%
%
If the strong \mgii\
systems arise in the centers of galaxies, then the cluster environment
does not affect them, so that their observed overdensity traces the
overdensity of cluster galaxies (with gas).   However, the weak \mgii\
population get destroyed in the cluster environment, and 
the fact that we do not detect an overdensity for them simply reflects that the
field contamination dominates in our redshift path.


Our observations allow us to put limits on the  shortage of
total cross-section for weak systems.  
%
%
 If the flat rEW distribution  is  due to truncated halos, the excess of
  galaxy counts in Table~\ref{tbl-galaxy} 
correspond to the missing fraction in absorbing cross section. We then
conclude that there is between one and two orders of magnitude less total
cross-section of \mgii\ gas having $W_0<0.3$\AA.

\subsection{Clustering}

Several studies have shown that strong ($W_0\ga1.0$) \mgii\ absorbers
trace overdense regions.  For example, Cooke et al. (2006) found that
damped \lya\ (DLA) systems cluster like LBGs, which themselves have a
non-negligible clustering signal. DLA systems also cluster around
quasars (Ellison et al. 2002; Russell, Ellison \& Benn 2006;
Prochaska, Hennawi \& Herbert-Fort 2007), just as galaxies cluster
around QSOs, and also around themselves possibly revealing large-scale
structure (Lopez \& Ellison 2003; Ellison \& Lopez 2002). More
recently, Bouche et al. (2007) have found clustering of strong \mgii\
systems around luminous red galaxies (LRGs).

At first glance our interpretation of gas truncation affecting only weak
absorbers seems to go in the opossite direction of the results by Bouche et
al. (2007). These authors find that LRGs correlate more strongly with weak
\mgii\ systems ($W_0\la 1$ \AA) than with strong ($W_0\ga 2$ \AA)
systems. From their bias ratio they derive absorber masses, and find that
stronger systems occur in galaxies associated with less massive dark-matter
halos ($M\sim 10^{11} M_\sun$) than weaker systems ($M\sim 10^{12}
M_\sun$). The Bouche et al. (2007) data, however, reaches only $W_0=0.3$
  \AA, while with our technique we probe much deeper in rEW.  In fact, from
  our results it follows just the opposite, namely that the stronger systems
correlate more strongly with galaxies: strong systems in our sample show more
clustering with clusters than weak systems. This apparent contradiction
becomes even more evident if one considers that LRG should flag clusters. But,
as already stated, $W_0<0.3$ \AA\ systems, might occur in much less massive
(dwarf) galaxies that were not probed in that study. Certainly a natural
follow-up of the present study will be to indentify the absorbing galaxies
from the RCS images and look for $W_0-L$ correlations.

Finally, let us note that rEW is basically a measure of the velocity spread
(Ellison 2006). One possible contribution to the overdensity of strong systems
observed in clusters could be that cluster absorbers have larger spreads due
to galaxy interactions, which is much more probable than for the field
absorbers. Indeed this effect has been proposed for 'ultra-strong' absorbers
($W_0>2.7$ \AA; Nestor et al. 2007). In addition, a mild correlation between
absorber assymetries and rEW has been found in the field (Kacprzak et
al. 2007) that could be strenghtened in our sample due to galaxy interactions.
The few cases in our high-resolution sample {\it S1} that show resolved
systems separated by several 100 \kms, all are weak systems. On the other
hand, the few strong systems that are both in {\it S1} and {\it S2} do not
show particular kinematics when observed at high resolution (e.g.,
velocity spans of several 100 
\kms). Clearly, a larger sample of strong cluster absorption systems must be
analyzed at high spectral resolution.

\subsection{Limits on the  fraction of neutral gas in clusters} 

Regardless of what produces the observed overdensity of \mgii\ in
clusters, we can  put constraints on the contribution of the
absorbing gas to the budget of cold baryons in clusters.
With its low ionization potential of $15$ eV, \mgii\ is a good tracer of
neutral gas. Indeed, several surveys have shown that  $W_0>0.6$
\AA\ \mgii\ systems frequently occur in DLA and sub-DLA
systems, i.e., in gas that is predominantly neutral (Rao \& Turnshek 2000;
Rao, Turnshek \& Nestor 2006).  Since ionization corrections are negligible
and since column densities in excess of $10^{20.3}$ \icm\ (the definition
threshold of a DLA system) can be obtained easily in low-resolution spectra,
measurements of the incidence of DLA systems have led to robust estimates of
the cosmological mass density of neutral gas, $\Omega_{\rm DLA}$.
At low redshift ($z_{abs}<1.6$), where confirmation of the DLA troughs at
$\lambda=1215$ \AA\ requires space-based observations, Rao, Turnshek \& Nestor
(2006) have searched for DLA systems using \mgii\ (redshifted to optical
wavelengths) as a signpost.  By measuring \ion{H}{1} column 
densities directly, these authors have found that $\approx 50$\%\ of \mgii\
systems with $2<W_0^{2796}<3$ \AA\ are DLA systems (with average column
densities $\langle N($\ion{H}{1}$)\rangle=3.5\pm 0.7\times 10^{20}$
\icm). According to these surveys (see also Rao \& Turnshek 2000), the mass
density provided by DLA systems at $\langle z \rangle=0.5$ is similar to the
high redshift value, $\Omega_{\rm DLA}=1\times 10^{-3}$. For a universal
baryon density of $\Omega_b=0.044$ (Spergel et al. 2006), $2.3$ \% of the
baryons in the Universe at $z=0.5$ is in DLA systems (at $z=0$ this fraction
falls down to $1$\%; Zwaan et al. 2003).

%


An overabundance of strong \mgii\ systems  of $\approx 10$, as observed in
  our cluster sample, with a 50\% chance of being a DLA system implies  a
factor of $5$ more neutral gas than the cosmic average. However, assuming
overdensities (by mass) of over two orders of magnitude at the typical cluster
radii probed here, $r_{200}$, yields a tiny 0.1\% of the cluster baryons in
form of neutral gas. This small amount of neutral gas seems more consistent
with that in present-day groups according to \ion{H}{1} 21 cm surveys (e.g.,
Pisano et al. 2007; Zwaan et al. 2003; Sparks, Carollo, \& Macchetto 1997).
If, as argued for DLA systems (e.g., Wolfe et al. 2004), the neutral gas has
served as fuel for star formation, then the small fraction of neutral gas in
the RCS clusters probed here may be taken as evidence that star-formation
either occurred at much earlier epochs than probed here $\langle z
\rangle=0.6$ or it was suppressed by the cluster environment early in the
accretion stage.

\subsection{Speculations}

The flattening of the rEW distribution we observe in clusters represents a
qualitative difference with the field in terms of absorber populations. This
difference strongly suggests that it is the cluster environment that drives
the morphological evolution of cluster galaxies, and not the field population
accreted by the clusters. If not, clusters would be more efficient in
accreting strong absorbers, which seems unlikely. Instead, it is more likely
that galaxies giving rise to weak absorbers have lost gas due to the cluster
environement.

The differing rEW distribution we observe in clusters could be also partly due
to a mix of evolutionary and morphological effects. Studies using imaging
stacking have shown (Zibetti et al. 2007) that strong absorbers arise in
bluer, later-type galaxies and weaker systems in red passive galaxies. If this
holds in our sample, it also fits well with our finding of a flat
rEW-distribution, considering that early-type galaxies in clusters evolve less
rapidly than later-type ones (Dressler et al. 1997).

 As already stated, local cluster galaxies show a deficit of \hi\ as a
  function of distance to the cluster centers. Already at $d\sim 1$ Mpc,
  \hi\ disks do not exceed the optical radii (e.g., Bravo-Alfaro et al. 2000).
  If our sample includes the high-redshift counterparts to these galaxies, the
  lack of weak \mgii\ overdensity may indicate that the processes giving rise
  to the stripping of gas were already in place at $z\approx 0.6$.  On the
  other hand, the denser gas (including molecular gas; Vollmer et al. 2005)
  survives the passages through the cluster center.  Using the above argument
  again, this gas, more internal to the galaxies, may host the strong
  absorbers we believe track the galaxy overdensities.




\section{Outlook}

We believe the present work opens a couple of important prospects,  both
  from the absorption-line and the host-galaxy perspectives.  First, the
high-resolution data can be used to perform further tests for the cluster
environment.  Are the ionization conditions the same as in field
\mgii\ systems? Does the kinematics of strong absorbers give any hint of
galaxy-galaxy interactions? Indeed, higher-ionization species such as
\civ\ and \ovi\ would perhaps be better suited for such tests (Mulchaey 1996),
but they require space-based observations.   Secondly, the galaxies giving
  rise to the observed \mgii\ in clusters must be identified and their
  properties compared with the field. Such a comparison should give important
  clues about the location of field \mgii\ absorbers.

Our experiment can be repeated with RCS-2, which will provide $10\times$ more
clusters, and also better photometric redshifts. With a larger sample one
could study possible evolutionary effects. For instance, is there an
absorption-line equivalent of the Butcher-Oemler effect? And, last but not
least, the 
role of gravitational lensing must be further explored, specially its possible
effect on the quasar luminosity function of cluster-selected samples.

%
%
%
%
%
%
%

\acknowledgements

We would like to thank Jason X. Prochaska and Sara L. Ellison for important
comments made on an earlier version of this paper. SL, LFB, PL and NP were
partly supported by the Chilean {\sl Centro de Astrof\'\i sica} FONDAP
No. 15010003. SL was also supported by FONDECYT grant N$^{\rm o}1060823$, and
LFB by FONDECYT grant N$^{\rm o}1040423$. The RCS project is supported by
grants to HY from the National Science and Engineering Research Council of
Canada and the Canada Research Chair Program. This research has made use of
the NASA/IPAC Extragalactic Database (NED) which is operated by the Jet
Propulsion Laboratory, California Institute of Technology, under contract with
the National Aeronautics and Space Administration.  Funding for the SDSS and
SDSS-II has been provided by the Alfred P. Sloan Foundation, the Participating
Institutions, the National Science Foundation, the U.S. Department of Energy,
the National Aeronautics and Space Administration, the Japanese
Monbukagakusho, the Max Planck Society, and the Higher Education Funding
Council for England. The SDSS Web Site is http://www.sdss.org/.


\clearpage

\begin{deluxetable}{lcccc}
\tablewidth{5.30in}
\tablecaption{High-resolution spectroscopic  quasar
  observations.\label{tbl-obs}}  
\tablehead{
\colhead{Quasar } & 
\colhead{$g$-mag} & 
\colhead{Exposure Time} &  
\colhead{S/N\tablenotemark{a}} &
\colhead{Date} 
}
\startdata
CTQ414               &17.0&  4500& 19   & Sept. 29 2006\\
022157.81+000042.5   &18.7&  7200& 19   & Sept. 24 2006\\
022239.83+000022.5   &18.5&  7200& 20   & Sept. 23 2006\\
022300.41+005250.0   &18.7&  4500& 24   & Sept. 24 2006\\
022441.09+001547.9   &18.9&  7200& 15   & Sept. 29 2006\\
022553.59+005130.9   &19.1&  3400& 10   & Sept. 30 2006\\
022839.32+004623.0   &19.0&  9900& 10   & Sept. 30 2006\\
CXOMP J054242.5-40   &18.9& 16200& 13   & March 18,19 2006\\
RXJ0911              &18.8& 43200& 51   & UVES Archive\\
Q1120+0195(UM425)    &15.7& 12900& 107  & March 18,19 2006\\
4974A\tablenotemark{b}&19.3& 11600&  8   & Sept. 23, 24 2006\\
HE2149-2745A         &16.8&  5400&  35  &   Sept. 29 2006\\
0918A\tablenotemark{b}&18.2&  7200&  15  &   Sept. 23 2006\\
231500.81-001831.2   &18.9&  7200&  18  &   Sept. 29 2006\\
231509.34+001026.2   &17.7&  7200&  33  &   Sept. 24 2006\\
231658.64+004028.7   &18.7&  7200&  13  &   Sept. 29 2006\\
231759.63-000733.2   &19.2&  7200&  10  &   Sept. 24 2006\\
231958.70-002449.3   &18.6&  7200&  15  &   Sept. 23 2006\\
232030.97-004039.2   &18.9&  9000&   8  &  Sept. 30 2006\\
\enddata
\tablenotetext{a}{Median signal-to-noise per pixel.}
\tablenotetext{b}{Newly discovered quasars. Named after Chandra fields.}
\end{deluxetable}

\clearpage

\begin{deluxetable}{clccccc}
\tabletypesize{\scriptsize}
\tablewidth{4.5in}
\tablecaption{\mgii\  Systems.\label{tbl-qsos}} 
\tablehead{
\colhead{LOS} & 
\colhead{Quasar} & 
\colhead{$z_{\rm em}$} & 
\colhead{$z_{\rm EW}$} & 
\colhead{$z_{\rm abs}$} & 
\colhead{$W_0^{2796}$ [\AA]} &
\colhead{$\sigma_{W_0^{2796}}$ [\AA]}\\  
\colhead{(1)} &
\colhead{(2)} &
\colhead{(3)} &
\colhead{(4)} &
\colhead{(5)} &
\colhead{(6)} &
\colhead{(7)} 
}
\startdata
  1 &CTQ414             & 1.29 &0.224 &0.3162& 0.484& 0.022\\
  2 &022157.81+000042.5 & 1.04 &0.237 &0.5919& 0.069& 0.008\\
    &                   &    * &    * &0.9812& 0.076& 0.009\\
    &                   &    * &    * &0.4190& 0.030& 0.009\\
  3 &022239.83+000022.5 & 0.99 &0.221 &0.6815& 0.695& 0.010\\
    &                   &    * &    * &0.8207& 0.118& 0.024\\
    &                   &    * &    * &0.7768& 0.122& 0.008\\
    &                   &    * &    * &0.7746& 0.150& 0.008\\
  4 &022300.41+005250.0 & 1.25 &0.212 &0.9493& 0.043& 0.010\\
  5 &022441.09+001547.9 & 1.20 &0.224 &1.0554& 0.881& 0.036\\
    &                   &    * &    * &0.9395& 0.080& 0.020\\
    &                   &    * &    * &0.6146& 0.181& 0.016\\
    &                   &    * &    * &0.3785& 1.181& 0.043\\
    &                   &    * &    * &0.2503& 0.732& 0.037\\
  6 &022553.59+005130.9 & 1.82 &0.383 &1.2253& 0.177& 0.032\\
    &                   &    * &    * &1.0945& 1.685& 0.065\\
    &                   &    * &    * &0.7494& 0.159& 0.015\\
    &                   &    * &    * &0.6816& 0.333& 0.019\\
  7 &022839.32+004623.0 & 1.29 &0.340 &0.6542& 0.597& 0.016\\
  8 &CXOMP J054242.5-40 & 1.44 &0.353 &1.0160& 0.414& 0.055\\
  9 &RXJ0911.4+0551     & 2.80 &0.190 &0.7684& 0.020& 0.002\\
    &                   &    * &    * &0.7747& 0.033& 0.002\\
    &                   &    * &    * &0.9946& 0.052& 0.002\\
    &                   &    * &    * &1.2100& 0.126& 0.002\\
 10 &Q1120+0195(UM425)  & 1.47 &0.203 &0.2476& 0.540& 0.005\\
 11 &4974A              & 1.50 &0.415 &0.7320& 0.388& 0.027\\
    &                   &    * &    * &0.4527& 0.115& 0.023\\
 12 &HE2149-2745A       & 2.03 &0.211 &0.6008& 0.175& 0.006\\
    &                   &    * &    * &0.6028& 0.015& 0.004\\
    &                   &    * &    * &0.4460& 0.016& 0.005\\
    &                   &    * &    * &0.4086& 0.228& 0.008\\
    &                   &    * &    * &0.5139& 0.028& 0.003\\
    &                   &    * &    * &1.0184& 0.219& 0.013\\
 13 &0918A              & 1.94 &0.237 &1.6055& 0.661& 0.021\\
    &                   &    * &    * &1.6105& 0.050& 0.010\\
 14 &231500.81-001831.2 & 1.32 &0.225 &0.5068& 0.063& 0.009\\
    &                   &    * &    * &0.5040& 0.148& 0.009\\
 15 &231509.34+001026.2 & 0.85 &0.209 &0.4470& 1.758& 0.009\\
 16 &231658.64+004028.7 & 1.05 &0.286 &0.4142& 0.137& 0.015\\
 17 &231759.63-000733.2 & 1.15 &0.382 &0.6010& 0.109& 0.016\\
 18 &231958.70-002449.3 & 1.89 &0.228 &0.4154& 0.192& 0.021\\
    &                   &    * &    * &0.4067& 0.151& 0.017\\
    &                   &    * &    * &0.8460& 2.028& 0.024\\
 19 &232030.97-004039.2 & 1.72 &0.355 &0.6980& 0.313& 0.012\\
 20 &022505.06+001733.2 & 2.42 &0.350 &0.9710& 1.610& 0.100\\
 21 &092142.03+384316.1 & 2.34 &0.350 &0.4730& 1.760& 0.100\\
 22 &092216.62+384448.0 & 0.59 &0.350 &0.5880& 1.080& 0.100\\
 23 &092746.94+375612.2 & 1.31 &0.350 &0.7780& 2.370& 0.100\\
 24 &092850.88+373713.0 & 1.45 &0.350 &1.3310& 3.320& 0.100\\
 25 &131623.99-015834.9 & 3.00 &0.350 &1.3140& 1.170& 0.100\\
 26 &141604.55+541039.6 & 1.49 &0.350 &1.0310& 1.790& 0.100\\
 27 &141635.78+525649.4 & 1.38 &0.350 &0.6980& 2.640& 0.100\\
 28 &141738.54+534251.1 & 2.58 &0.350 &0.7280& 1.580& 0.100\\
 29 &141838.36+522359.3 & 1.12 &0.350 &1.0230& 1.470& 0.100\\
 30 &141905.17+522527.7 & 1.61 &0.350 &0.4920& 1.120& 0.100\\
 31 &142043.68+532206.3 & 1.72 &0.350 &0.7650& 1.570& 0.100\\
    &                   &    * &    * &1.6980& 1.900& 0.100\\
 32 &142106.86+533745.1 & 1.86 &0.350 &0.8510& 1.800& 0.100\\
 33 &231710.78+000859.0 & 1.68 &0.350 &1.7970& 2.140& 0.100\\
 34 &231912.83+002046.6 & 1.23 &0.350 &1.1400& 1.560& 0.100\\
 35 &232001.05-005450.5 & 1.69 &0.350 &1.4220& 2.440& 0.100\\
 36 &232007.52+002944.3 & 0.94 &0.350 &0.9090& 1.240& 0.100\\
 37 &232133.76-010645.  & 1.98 &0.350 &1.5200& 1.220& 0.100\\
 38 &232208.09+005948.3 & 1.47 &0.350 &1.1950& 2.480& 0.100\\
    &                   &    * &    * &1.4100& 2.520& 0.100\\
\enddata
\tablecomments{
Table columns: (1) Line-of-sight Numbering [LOS 1--19: Absorbers found in
  sample {\it S1}; LOS 20--31: Absorbers found in sample {\it S2}]; (2) Quasar
  Name; (3) Emission redshift; (4) Minimum redshift for  a $3\sigma$ detection
  of the \mgii\ $\lambda 2796$ line with  $W>W_0^{min}$; 
  (5) \mgii\ absorption redshift; (6) and (7) Rest-frame equivalent width of
  \mgii\ $\lambda 2796$ in
  \AA\  and $1\sigma$ error.
}  
\end{deluxetable}

\clearpage

\begin{deluxetable}{clccccr}
\tablewidth{4.8in}
\tabletypesize{\scriptsize}
\tablecaption{Galaxy Clusters.\label{tbl-clusters}} 
\tablehead{
\colhead{LOS} & 
\colhead{Cluster} & 
\colhead{$z_{\rm cluster}$} &  
\colhead{$z_{\rm min}$} &  
\colhead{$z_{\rm max}$} &  
\colhead{$d$ [arcmin]} & 
\colhead{$d$ [\hkpc]}\\
\colhead{(1)} &
\colhead{(2)} &
\colhead{(3)} &
\colhead{(4)} &
\colhead{(5)} &
\colhead{(6)} &
\colhead{(7)} 
}
\startdata
  1& group/cluster    &0.500\tablenotemark{a} &0.400 &0.600  &0.19 &  69.6\\
  2& RCS022200+0000.1 &0.270 &0.237 &0.370  &5.99 &1475.5\\
  3& RCS022239+0001.7 &0.502 &0.402 &0.602  &1.37 & 499.6\\
   & RCS022221+0001.1 &0.270 &0.221 &0.370  &4.59 &1130.9\\
  4& RCS022302+0052.9 &0.509 &0.409 &0.609  &0.56 & 208.1\\
   & RCS022253+0055.1 &0.939 &0.839 &1.039  &2.93 &1390.2\\
  5& RCS022443+0017.6 &0.431 &0.331 &0.531  &1.96 & 656.4\\
   & RCS022436+0014.2 &0.173 &0.224 &0.273  &1.88 & 328.2\\
   & RCS022431+0018.0 &0.480 &0.380 &0.580  &3.33 &1188.6\\
   & RCS022449+0016.2 &0.818 &0.718 &0.918  &2.07 & 941.3\\
   & RCS022454+0013.3 &0.511 &0.411 &0.611  &4.07 &1503.6\\
  6& RCS022546+0050.0 &0.873 &0.773 &0.973  &2.27 &1052.1\\
   & RCS022556+0052.7 &0.928 &0.828 &1.028  &1.48 & 700.6\\
   & RCS022553+0052.5 &0.423 &0.383 &0.523  &1.00 & 332.6\\
   & RCS022602+0055.5 &0.352 &0.383 &0.452  &4.57 &1349.7\\
   & RCS022558+0051.8 &0.701 &0.601 &0.801  &1.30 & 557.7\\
  7& RCS022828+0044.9 &1.032 &0.932 &1.132  &2.98 &1450.7\\
   & RCS022829+0045.8 &0.774 &0.674 &0.874  &2.51 &1118.7\\
   & RCS022832+0046.5 &0.629 &0.529 &0.729  &1.82 & 743.8\\
   & RCS022841+0044.9 &0.271 &0.340 &0.371  &1.50 & 371.7\\
   & RCS022844+0047.7 &0.516 &0.416 &0.616  &1.85 & 685.6\\
  8& 054240.1-405503  &0.634\tablenotemark{b} &0.624 &0.644  &3.70 &1519.8\\
   & [BGV2006] 015    &0.502\tablenotemark{b} &0.492 &0.512  &3.70 &1353.1\\
   & [BGV2006] 018    &0.527\tablenotemark{b} &0.517 &0.537  &2.90 &1088.4\\
  9& RX J0911+05      &0.769\tablenotemark{c} &0.759 &0.779  &0.70 & 311.3\\
 10& UM425            &0.770\tablenotemark{d} &0.760 &0.780  &0.10 &  44.5\\
 11& MS2137.3-2353    &0.313\tablenotemark{e} &0.303 &0.323  &1.95 & 532.6\\
 12& group/cluster    &0.700\tablenotemark{a} &0.600 &0.800  &0.18 &  77.2\\
 13& CLJ2302.8+0844   &0.722\tablenotemark{f} &0.712 &0.732  &1.50 & 651.1\\
 14& RCS231515-0015.6 &0.566 &0.466 &0.666  &4.58 &1782.4\\
   & RCS231506-0018.1 &0.560 &0.460 &0.660  &1.59 & 614.2\\
   & RCS231501-0013.6 &0.557 &0.457 &0.657  &4.92 &1901.2\\
   & RCS231515-0015.8 &0.496 &0.396 &0.596  &4.49 &1631.9\\
   & RCS231459-0018.9 &0.522 &0.422 &0.622  &0.52 & 195.3\\
   & RCS231512-0020.1 &0.517 &0.417 &0.617  &3.43 &1275.2\\
 15& RCS231509+0012.1 &0.420 &0.320 &0.520  &1.77 & 583.3\\
 16& RCS231725+0036.6 &0.266 &0.286 &0.366  &7.70 &1876.6\\
 17& RCS231755-0011.3 &0.573 &0.473 &0.673  &3.92 &1536.7\\
 18& RCS231947-0028.3 &0.651 &0.551 &0.751  &4.52 &1879.1\\
   & RCS231944-0027.0 &0.805 &0.705 &0.905  &4.28 &1934.4\\
   & RCS231944-0026.8 &0.844 &0.744 &0.944  &4.01 &1840.3\\
   & RCS231958-0023.2 &0.796 &0.696 &0.896  &1.59 & 716.9\\
   & RCS231958-0025.1 &0.789 &0.689 &0.889  &0.31 & 141.1\\
 19& RCS232028-0043.0 &1.085 &0.985 &1.185  &2.51 &1234.6\\
   & RCS232029-0038.1 &0.589 &0.489 &0.689  &2.50 & 990.7\\
   & RCS232027-0042.7 &0.853 &0.753 &0.953  &2.30 &1060.0\\
 20& RCS022443+0017.6 &0.431 &0.350 &0.531  &5.43 &1820.9\\
   & RCS022527+0015.2 &0.345 &0.350 &0.445  &6.07 &1770.4\\
   & RCS022454+0013.3 &0.511 &0.411 &0.611  &5.01 &1850.0\\
   & RCS022449+0016.2 &0.818 &0.718 &0.918  &4.18 &1902.4\\
 21& RCS092148+3841.2 &0.961 &0.861 &1.061  &2.35 &1123.3\\
   & RCS092130+3843.8 &0.373 &0.350 &0.473  &2.33 & 714.1\\
   & RCS092123+3836.3 &0.285 &0.350 &0.385  &7.82 &1999.0\\
   & RCS092131+3845.6 &0.519 &0.419 &0.619  &3.13 &1165.5\\
 22& RCS092223+3842.0 &0.501 &0.401 &0.590  &3.05 &1115.5\\
   & RCS092219+3846.1 &0.598 &0.498 &0.590  &1.40 & 559.5\\
   & RCS092222+3841.4 &0.434 &0.350 &0.534  &3.57 &1201.1\\
 23& RCS092809+3754.8 &0.441 &0.350 &0.541  &4.73 &1607.4\\
   & RCS092753+3755.5 &0.844 &0.744 &0.944  &1.41 & 647.6\\
   & RCS092811+3756.3 &0.569 &0.469 &0.669  &4.82 &1881.5\\
   & RCS092743+3800.2 &0.540 &0.440 &0.640  &4.08 &1548.8\\
   & RCS092742+3759.2 &0.875 &0.775 &0.975  &3.22 &1495.2\\
   & RCS092731+3754.3 &0.768 &0.668 &0.868  &3.61 &1605.9\\
   & RCS092740+3756.9 &0.773 &0.673 &0.873  &1.49 & 665.0\\
 24& RCS092901+3738.6 &0.882 &0.782 &0.982  &2.48 &1157.5\\
   & RCS092843+3741.7 &0.368 &0.350 &0.468  &4.74 &1440.7\\
   & RCS092905+3739.5 &0.640 &0.540 &0.740  &3.65 &1507.0\\
   & RCS092843+3734.6 &0.755 &0.655 &0.855  &2.93 &1295.9\\
   & RCS092851+3733.7 &0.769 &0.669 &0.869  &3.43 &1525.7\\
 25& RCS131627-0157.6 &0.897 &0.797 &0.997  &1.18 & 550.8\\
 26& RCS141601+5410.4 &1.387 &1.287 &1.487  &0.48 & 244.6\\
   & RCS141617+5407.3 &0.943 &0.843 &1.043  &3.78 &1795.4\\
   & RCS141546+5407.8 &0.698 &0.598 &0.798  &3.87 &1658.0\\
 27& RCS141627+5256.8 &0.687 &0.587 &0.787  &1.19 & 504.6\\
 28& RCS141749+5341.4 &1.384 &1.284 &1.484  &2.18 &1112.3\\
   & RCS141756+5344.3 &0.378 &0.350 &0.478  &3.06 & 946.6\\
   & RCS141724+5342.7 &0.649 &0.549 &0.749  &2.04 & 844.8\\
 29& RCS141803+5223.1 &0.271 &0.350 &0.371  &5.40 &1333.4\\
   & RCS141838+5225.7 &0.614 &0.514 &0.714  &1.72 & 695.8\\
   & RCS141824+5228.0 &0.328 &0.350 &0.428  &4.52 &1273.7\\
   & RCS141840+5221.8 &0.838 &0.738 &0.938  &2.22 &1018.8\\
 30& RCS141946+5223.3 &0.337 &0.350 &0.437  &6.72 &1930.6\\
   & RCS141923+5228.5 &0.400 &0.350 &0.500  &4.11 &1317.9\\
   & RCS141838+5225.7 &0.614 &0.514 &0.714  &4.09 &1653.9\\
   & RCS141824+5228.0 &0.328 &0.350 &0.428  &6.65 &1873.0\\
 31& RCS142018+5322.3 &1.147 &1.047 &1.247  &3.70 &1841.2\\
 32& RCS142111+5339.8 &0.825 &0.725 &0.925  &2.23 &1014.9\\
 33& RCS231711+0012.5 &0.521 &0.421 &0.621  &3.51 &1309.6\\
 34& RCS231924+0023.1 &0.908 &0.808 &1.008  &3.64 &1713.1\\
   & RCS231923+0020.5 &0.285 &0.350 &0.385  &2.58 & 658.9\\
 35& RCS232013-0053.2 &0.835 &0.735 &0.935  &3.54 &1619.9\\
   & RCS232016-0056.7 &0.695 &0.595 &0.795  &4.23 &1807.4\\
 36& RCS231952+0028.4 &0.569 &0.469 &0.669  &3.94 &1539.8\\
 37& RCS232148-0104.1 &0.732 &0.632 &0.832  &4.57 &1994.4\\
 38& RCS232158+0100.3 &0.684 &0.584 &0.784  &2.58 &1096.0\\
   & RCS232157+0100.2 &0.856 &0.756 &0.956  &2.57 &1187.7\\
   & RCS232206+0101.6 &0.661 &0.561 &0.761  &1.86 & 780.1\\
%
                                                         
\enddata                                                 
\tablecomments{Table displays only lines-of-sights with detected absorption.}
\tablecomments{Table columns: (1) Line-of-sight Numbering (same as in
  Table~\ref{tbl-qsos}); (2) Cluster Name;               
  (3) Cluster Redshift [references other than the RCS are: $^a$Faure et
  al. (2004), $^b$Barkhouse et al. (2006), $^c$Kneib, Cohen \& Hjorth (2000),
  $^d$Green et al. (2005), $^e$Stocke et al. (1991), $^f$Perlman et
  al. (2002)]; (4) and (5) Minimum and Maximum redshift surveyed,
  respectively; (6) and (7) Projected distance in arcminutes 
and  physical distance at \zc, respectively,  from quasar
  line-of-sight to   cluster   coordinates. 
}  
\end{deluxetable}

\clearpage

\begin{table}
\caption{Statistical Samples.\label{tbl-samples}}
\begin{tabular}{lcccccccccccc}
\tableline\tableline
&&& Quasars&&\multicolumn{2}{c}{Clusters}&&\multicolumn{2}{c}{Pairs}&&\multicolumn{2}{c}{Absorbers}\\
\cline{4-4} \cline{6-7} \cline{9-10}\cline{12-13}\\
    &$z_{\rm min}$ & $z_{\rm max}$    & \# && \#\tablenotemark{a} &$\langle B_{gc}\rangle$ &&\#\tablenotemark{a} &$\Delta$\zc&& $W_0^{\rm min}$[\AA]  &
\#\tablenotemark{b}\\  
\tableline
%
SDSS-RCS     & 0.20 &  ...  &     190&&    368 & ... && 442 &    ...&&     ...  &...\\
{\it S1     }& 0.20 &  0.90 &      19&&     46 & 327 &&  46 &   6.32&&      0.05&  37\\
{\it S2     }& 0.35 &  0.90 &     144&&    255 & 263 && 375 &  57.01&&    1.0   &  23\\
{\it S2-best}& 0.35 &  0.90 &      88&&    104 & 488 && 125 &  18.06&&    1.0   &  14\\
%
%
\tableline
\end{tabular}
\tablenotetext{}{Note.---These samples are not disjoint.}
\tablenotetext{a}{Number of objects having $z_{\rm min}-\delta z<$ \zc $<
  z_{\rm max}+\delta z$}
\tablenotetext{b}{Total number of systems with $W>W_0^{\rm min}$}
\end{table}

\clearpage

\begin{table}
\caption{Redshift Path Density of \mgii\ in Clusters at $\langle z \rangle=0.6$.\label{tbl-hits}}
\begin{tabular}{lcccccc}
\tableline\tableline
Sample & $\Delta$\zc\tablenotemark{a} &  $W_0^{2796}$ [\AA] & $N_{\rm hits}$\tablenotemark{a} &
 $(dN/dz)_c$\tablenotemark{b}& $(dN/dz)_f$\tablenotemark{c}& Overdensity $\delta$   \\ 
\tableline
\multicolumn{7}{c}{$d<2$ \hmpc}\\
\cline{1-7} \\
{\it S1 }&    6.32 & $[0.05,0.3]$&          5 &       0.79(0.31,1.67)    &     1.09  &   0.7\\
{\it S1 }&    6.32 & $>0.3$      &          6 &       0.95(0.41,1.88)    &     0.68  &   1.4\\
{\it S1 }&    6.32 & $>0.6$      &          4 &       0.63(0.22,1.45)    &     0.42  &   1.5\\
{\it S2 }&   57.01 & $>1.0$      &          9 &       0.16(0.09,0.29)    &     0.16  &   1.0\\
{\it S2 }&   57.01 & $>2.0$      &          3 &       0.05(0.02,0.14)    &     0.040 &   1.3\\
{\it S2 }&   57.01 & $[2.0,3.0]$ &          3 &      0.053(0.015,0.141)  &0.033\tablenotemark{d} &   1.6\tablenotemark{d}\\ 
{\it S2-best}&18.06&  $>1.0$     &          5 &         0.28(0.11,0.58)  &    0.16   &  1.8\\
{\it S2-best}&18.06&  $>2.0$     &          2 &         0.11(0.02,0.35)  &    0.040  &  2.8\\
\tableline
\multicolumn{7}{c}{$d<1$ \hmpc}\\
\cline{1-7} \\
{\it S1} &   3.33  & $[0.05,0.3]$&        4  &       1.20(0.41,2.75)   &    1.09   &  1.1 \\
{\it S1} &   3.33  & $>0.3$      &        6  &       1.80(0.75,3.51)   &    0.68   &  2.6 \\
{\it S1} &   3.33  & $>0.6$      &        4  &       1.20(0.41,2.75)   &    0.42   &  2.9 \\
{\it S2} &  14.13  & $>1.0$      &        7  &       0.50(0.23,0.93)   &    0.16   &  3.1 \\
{\it S2} &  14.13  & $>2.0$      &        3  &       0.21(0.06,0.55)   &    0.040  &  5.3 \\
{\it S2} &  14.13  & $[2.0,3.0]$ &        3  &      0.212(0.058,0.549) &    0.033\tablenotemark{d}  &  6.4\tablenotemark{d} \\
{\it S2-best}& 5.51& $>1.0$      &        4  &        0.73(0.25,1.66)  &    0.16   &  4.5\\
{\it S2-best}& 5.51& $>2.0$      &        2  &        0.36(0.06,1.14)  &    0.040  &  9.1\\
\tableline
\multicolumn{7}{c}{$d<0.5$ \hmpc}\\
\cline{1-7} \\
{\it S1} &  1.45   & $[0.05,0.3]$&       3   &     2.07(0.57,5.35)     &   1.09    & 1.9 \\
{\it S1} &  1.45   & $>0.3$      &       2   &     1.38(0.25,4.35)     &   0.68    & 2.0 \\
{\it S1} &  1.45   & $>0.6$      &       2   &     1.38(0.25,4.35)     &   0.42    & 3.3 \\
{\it S2} &  3.72   & $>1.0$      &       1   &     0.27(0.01,1.28)     &   0.16    & 1.7 \\
{\it S2} &  3.72   & $>2.0$      &       1   &     0.27(0.01,1.28)     &   0.040   & 6.8 \\
{\it S2} &  3.72   & $[2.0,3.0]$ &       1   &    0.269(0.014,1.275)   &   0.033\tablenotemark{d}   & 8.2\tablenotemark{d} \\
{\it S2-best}&  0.79&  $>1.0$    &         0 & &&\\  
\tableline
\end{tabular}
\tablenotetext{a}{Between $z=z_{\rm min}$ and $z=0.9$}
\tablenotetext{b}{Cluster redshift density with  $95$\%  confidence limits}
\tablenotetext{c}{
Field redshift density. $W_0>1$ \AA\ cut  from
PPB06; $W_0>2$ and $W_0>0.6$ \AA\ cuts from NTR05 using
$(dN/dz)_f=1.001(1+z)^{0.226}\exp[{-(W_0/0.443)(1+z)^{-0.634}}]$; and
$W_0<0.3$ \AA\ cuts  from Churchill et al. (1999) with
$(dN/dz)_f=0.8(1+z)^{1.3}$ and a 76.7\%\ downward correction due to their
smaller $W_0^{\rm min}=0.02$ \AA.}
\tablenotetext{d}{NTR06 find $(dN/dz)_f \approx 0.015$, implying a factor of $\sim
  2.2$ 
  higher overdensity in this bin at the $>3\sigma$ level.}
\end{table}

\begin{deluxetable}{cccc}
\tablewidth{5in}
\tablecaption{Expected Galaxy Overdensity.\label{tbl-galaxy}} 
\tablehead{
\colhead{$\log_{10}(M/M_{\sun})$} & 
\colhead{$d<2$ \hmpc} & 
\colhead{$d<1$ \hmpc} &
\colhead{$d<0.5$ \hmpc} \\
}
\startdata

13 & 1.7   &  8.2 &34.0\\
14 & 10.0  & 40.0 &132.0\\
\enddata
\end{deluxetable}

\clearpage

\clearpage

\clearpage

\begin{figure}
\epsscale{.80}
\includegraphics[angle=0,scale=0.8]{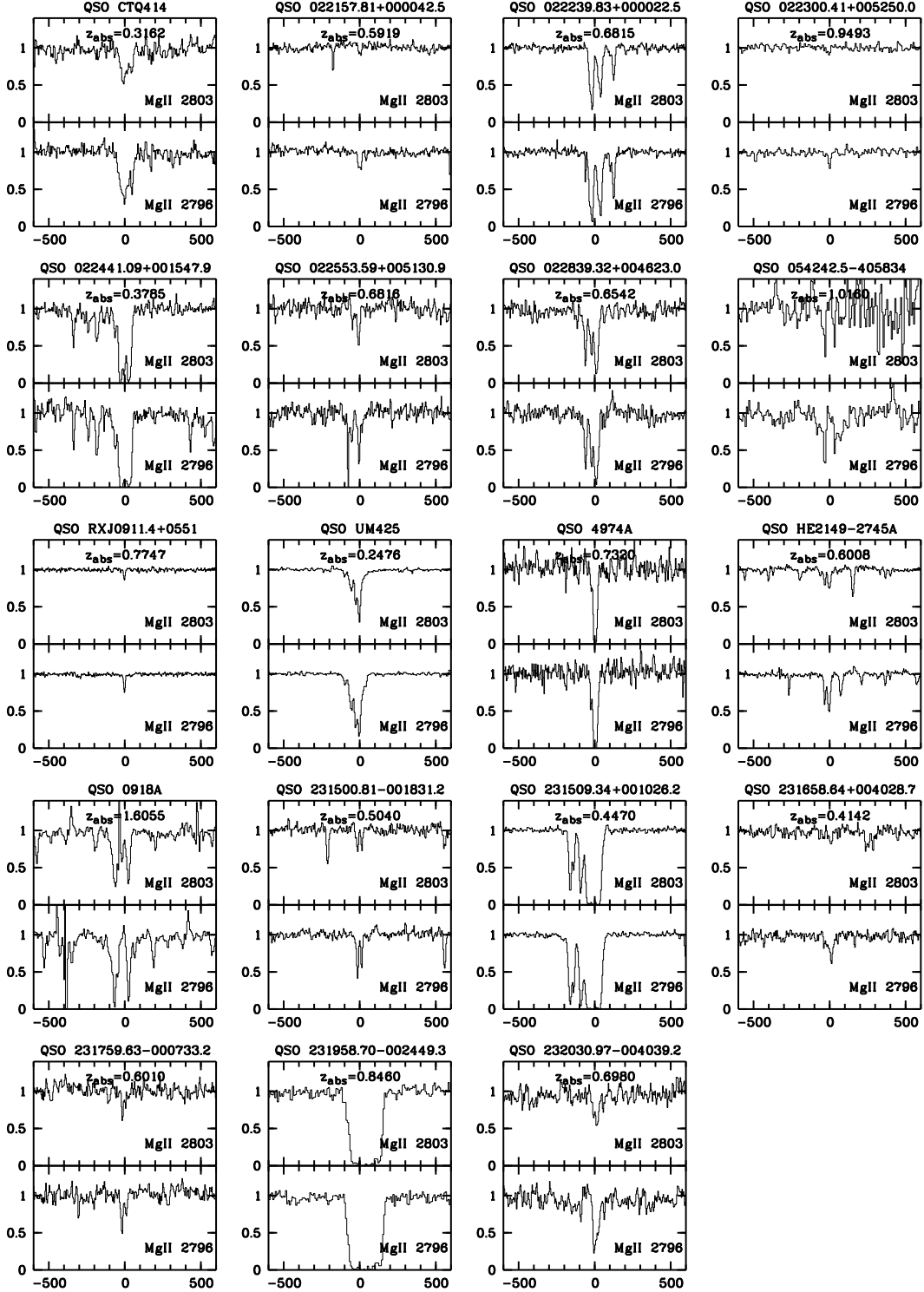}
\caption{ Selected \mgii\ absorption line systems in each of the 19 spectra
  comprising the high-resolution sample, {\it S1}.  Each panel (normalized
  flux vs. rest-frame velocity in \kms) shows the strongest \mgii\ doublet in
  the spectrum, unless an absorption redshift is within [$z_{\rm min},z_{\rm
  max}$] of a cluster in the same LOS, in which case that latter system is
  plotted.  Associated systems ($z_{\rm abs}\sim z_{\rm em}$) were not
  considered\label{fig_spec}}
\end{figure}

\clearpage

\begin{figure}
\epsscale{.80}
\includegraphics[angle=-90,scale=0.8]{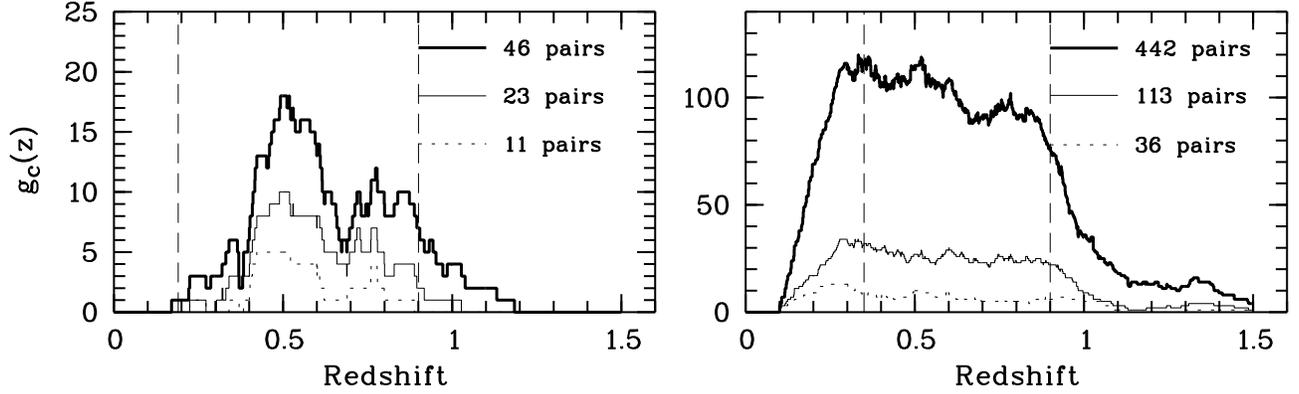}
\caption{ Cluster redshift-path density, $g_c(W_0^{\rm min},z_i)$ of the
high-resolution sample ({\it S1}, $W_0^{\rm min}=0.05$ \AA, lefthand panel)
and low-resolution sample ({\it S2}, $W_0^{\rm min}=1.0$ \AA). The thick line
is for LOS-cluster distances $d<2$, the thin line for $d<1$, and the dotted
line for $d<0.5$ \hmpc. The vertical dashed lines depict the redshift defined
by the rEW detection thresholds.  See \S~\ref{sect-g} for more
details.\label{fig_g}}
\end{figure}

\clearpage

\begin{figure}
\epsscale{.80}
\includegraphics[angle=0,scale=1]{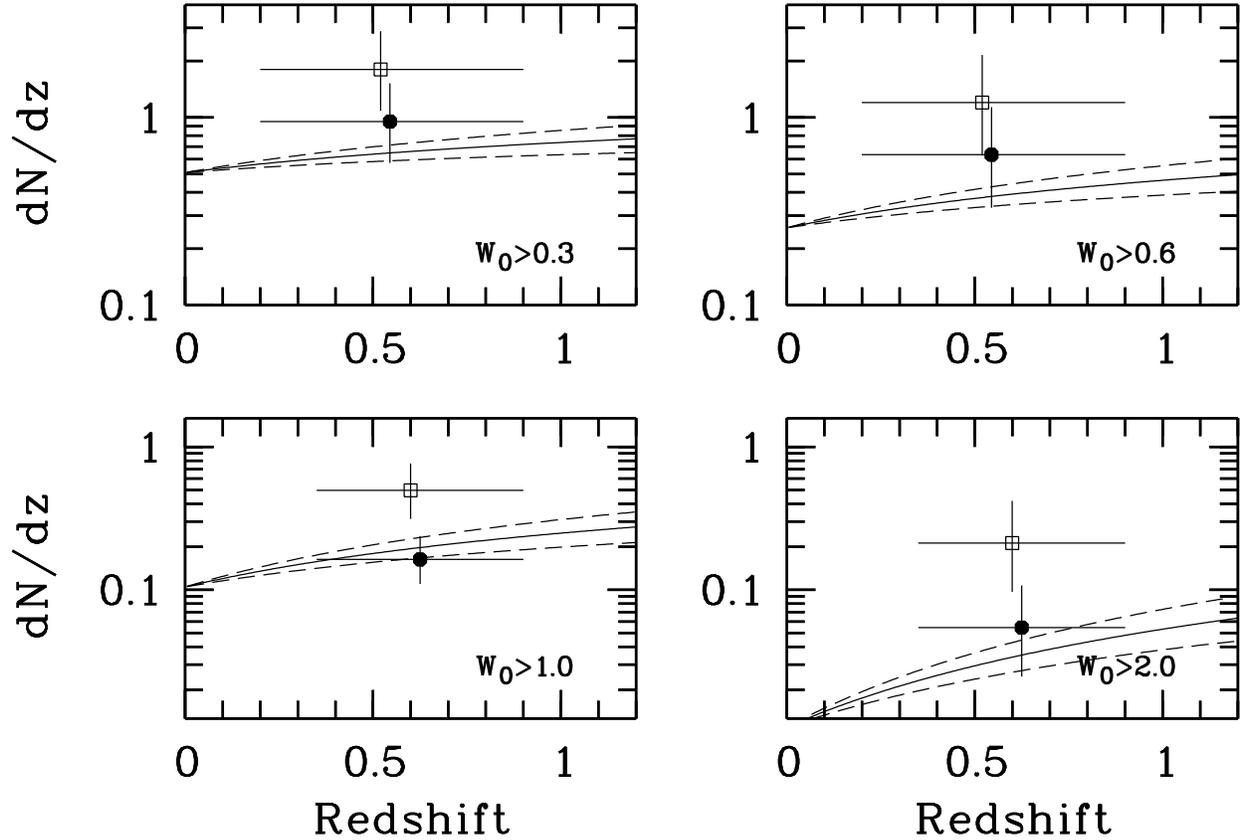}
\caption{ \mgii\ redshift number density binned in the entire range of
  cluster redshifts for various $W_0^{2796}$ lower limits. The filled
  circles are from clusters with LOS-cluster distances $d<2$ \hmpc\
  and the open squares from clusters with $d<1$ \hmpc\ (symbols
  slightly shifted in the x-axis for more clarity).  The errors bars
  correspond to $1\sigma$. The curves correspond to the fit by NTR05
  to their SDSS EDR data of field absorbers along with $1\sigma$
  limits.  The top panels show results from sample {\it S1} only
  (high-resolution spectra; 46 pairs, \zave$=0.550$), while points in
  the bottom panels were calculated using only the {\it S2} sample
  (375 pairs; \zave$=0.625$).\label{fig_n_z}}

\end{figure}


%
%
\clearpage

\begin{figure}
\epsscale{.80}
\includegraphics[angle=-90,scale=0.7]{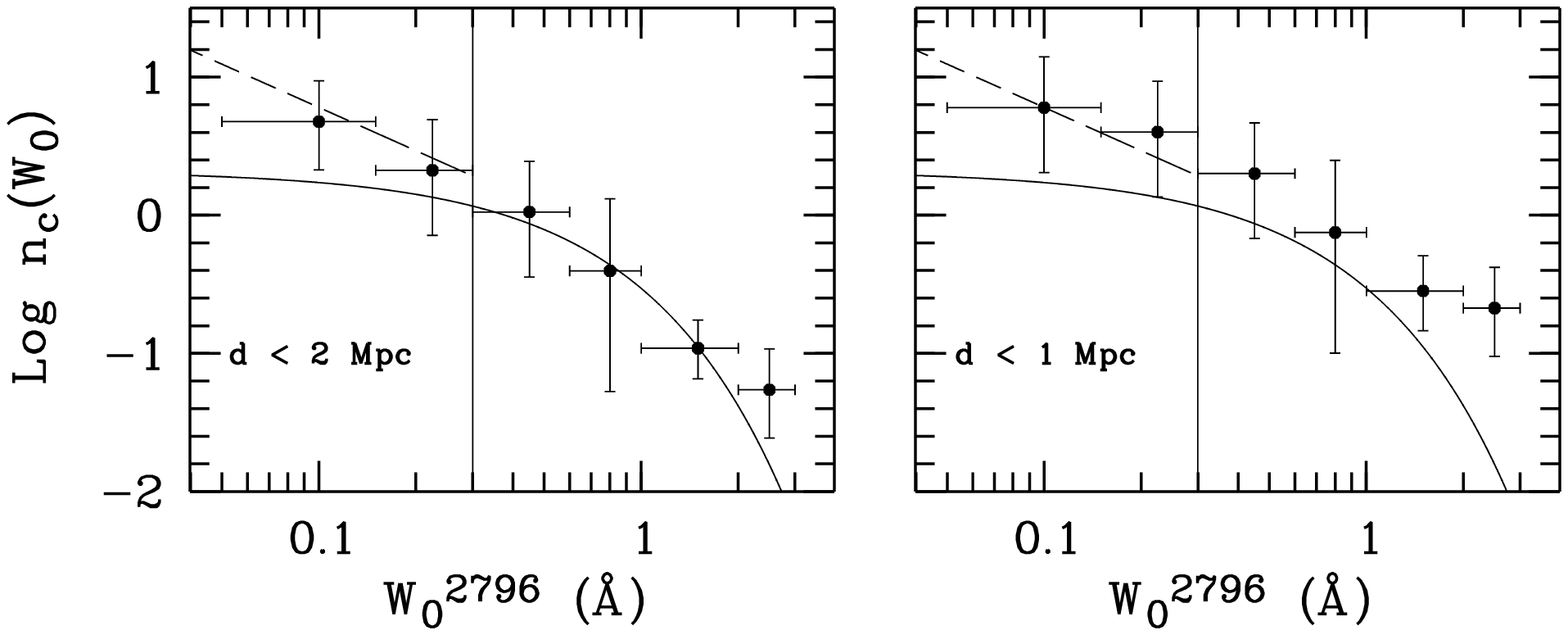}
\caption{ Equivalent width distribution of \mgii\ absorbers in clusters
  (corrected by the cluster redshift path) vs. \mgii\ $\lambda 2796$
  rest-frame equivalent width for LOS-cluster distances $d<2$ \hmpc\ (lefthand
  panel) and $d<1$ \hmpc.  The errors bars correspond to $1\sigma$. Data
  points with $W_0<1.0$ \AA\ resulted from sample {\it S1} only
  (high-resolution spectra), while points at $W_0>1.0$ \AA\ resulted from
  sample {\it S2}.  The lines are the field expectations. The solid line is
  the exponential distribution fitted by NTR06 to their MMT data having
  $W_0>0.3$ \AA, and the dashed curve is the power-law fited by CRCV99
 to their HIRES data having $W_0<0.3$ \AA. The vertical line at
  $W_0=0.3$ \AA\ marks the transition in $n(W)$ pointed out by
  NTR06. \label{fig_n_W}}
\end{figure}

%
%


\begin{thebibliography}{}
\bibitem[]{} Barkhouse, W. A., Green, P. J., Vikhlinin, A., et al. 2006, ApJ, 645, 955
\bibitem[]{} 
Bartelmann, M., \& Loeb, A. 1996, ApJ, 457, 529
\bibitem[]{} 
Bergeron, J., \& Stasinska, G. 1986, A\&A, 169, 1
\bibitem[]{} 
Bergeron, J., \& Boiss\'e, P. 1991, A\&A, 243, 344
\bibitem[]{} 
Bouche, N.,  Murphy, M. T., Peroux, C.,  Csabai, I., \& Wild, V.  2006, MNRAS,
371, 495 (BMPCW06)
\bibitem[]{} 
Boyle, B. J., Fong, R., \& Shanks, T. 1988, MNRAS, 231, 897
\bibitem[]{} 
Bravo-Alfaro, H., Cayatte, V., van Gorkom, J. H., \& Balkowski, C. 2000, AJ,
119, 580 
\bibitem[]{} 
Butcher, H. \& Oemler, G. 1984, ApJ, 285, 426
\bibitem[]{} 
Caulet, A. 1989, ApJ, 340, 90
\bibitem[]{}
Chung, A.,  van Gorkom, J. H., Kenney, J. D. P. \& Vollmer, B. 
2007, ApJ, 659, 115 
\bibitem[]{}
Churchill, C. W., Rigby, J. R., Charlton, J. C., \& Vogt, S. S. 1999, ApJS,
120, 51 (CRCV99)
\bibitem[]{} 
Churchill, C.W., Mellon, R.R., Charlton, J.C., Jannuzi, B.T., Kirhakos, S.,
Steidel, C.C., \& Schneider, D.P. 2000, ApJ, 543, 577 
\bibitem[]{} 
Churchill, C. W., Vogt, S. S., \& Charlton, J. C. 2003, ApJ, 125, 98
\bibitem[]{} 
Churchill et al. 2005, IAU Conference 199, Shangai
\bibitem[]{} 
Churchill, C. W., Kacprzak, G. G., Steidel, C. C., \& Evans, J. L. 2007, ApJ,
arXiv:astro-ph/0612560  
\bibitem[]{}
Cooke, J., Wolfe, A. M., Gawiser, E. \& Prochaska, J. X. 2006, ApJ,
636, 9	
\bibitem[]{} 
Cooray, A, 2006, MNRAS, 365, 842
\bibitem[]{} 
Croton, D., et al., 2006, MNRAS, 365, 11
\bibitem[]{} 
Dressler, A. 1980, ApJ, 236, 351
\bibitem[]{}
Ellison, S. L. \&  Lopez, S., 2001, A\& A, 380, 117
\bibitem[]{}
Ellison, S. L., Yan, L., Hook, I. M., Pettini, M., Wall, J. V. \&
Shaver, P. 2002, A\&A  383, 91
\bibitem[]{} 
Ellison, S. L., Churchill, C. W., Rix, S. A., \& Pettini, M. 2004a, ApJ,
615, 118
\bibitem[]{}
Ellison, S. L., Ibata, R., Pettini, M., Lewis, G. F., Aracil, B.,
Petitjean, P. \&  Srianand, R. 2004b A\&A, 414, 79
\bibitem[]{} 
Ellison, S. L., Kewley, L. J., \&  Mall\'en-Ornelas, G., 2005, MNRAS,
357, 354
\bibitem[]{} 
Ellison S. L., 2006, MNRAS, 368, 335
\bibitem[]{} 
Ettori, S. 2003, MNRAS, 344, L13
\bibitem[]{} 
Fassnacht, C. D., et al. 2006, ApJ, 651, 667
\bibitem[]{} 
Faure, C., Alloin, D., Kneib, J. P., \& Courbin, F., 2004, A\&A, 428, 741 
\bibitem[]{} 
Gehrels, N., 1986, ApJ, 303, 336
\bibitem[]{} 
Gilbank, D., Yee, H.~K.~C., Ellingson, E., Gladders, M. D., Barrientos,
L.~F. \&  Blindert, K.\ 2007, \aj, 134, 282 
\bibitem[]{}
Giovanelli, R. \& Haynes, M. P. 1983, AJ, 88, 881
\bibitem[]{} Gladders, M.~D., Yee, 
H.~K.~C., Majumdar, S., Barrientos, L.~F., Hoekstra, H., Hall, P.~B., \& 
Infante, L.\ 2007, \apj, 655, 128 
\bibitem[Gladders \& Yee(2005)]{gladders2005} Gladders, M.~D., \& 
Yee, H.~K.~C.\ 2005, \apjs, 157, 1 
\bibitem[Gladders \& Yee(2000)]{gladders2000} Gladders, M.~D., \& 
Yee, H.~K.~C.\ 2000, \aj, 120, 2148 
\bibitem[]{} 
Green, P. J., Infante, L., Lopez, S., Aldcroft, T. L., \& Winn, J. N. 2005,
ApJ, 630, 142
\bibitem[]{} 
Kacprzak, G. G., Churchill, C. W., Steidel, C. C., Murphy, M. T., \& Evans,
J. L. 2007, ApJ, 662, 909
\bibitem[]{} 
Kneib, J.-P., Cohen, J. G., \& Hjorth, J. 2000, ApJ, 544, L35
\bibitem[]{} 
Koester, B.~P. et al. 2007, \apj, 660, 239 
\bibitem[]{}
Lanzetta, K.M., Turnshek, D.A., \& Wolfe, A.M. 1987, ApJ, 322, 739
\bibitem[]{} 
Lanzetta, K.M., \& Bowen, D. 1990, ApJ, 357, 321
\bibitem[]{} 
Le Brun, V., Bergeron, J., Boisse, P., \& Deharveng, J. M. 2001, A\&A, 321, 733
\bibitem[]{} 
Lopez, S. \&  Ellison, S. L., 2003, A\& A, 403, 573
\bibitem[]{} 
Lynch, R. S., Charlton, J. C., \& Kim, T. S. 2006, ApJ, 640, 81
\bibitem[]{} 
Maller A. H. \&  Bullock J. S., 2004, MNRAS, 355, 694
McCarthy, I. G., Bower, R. G., \& Balogh, M. L. 2007, MNRAS,
arXiv:astro-ph/0609314 
\bibitem[]{} 
M\'enard, B., Nestor, D., Turnshek, D., Quider, A., Richards, G., Chelouche,
D., \& Rao, S. 2007, ApJ (arXiv:0706.0898) 
\bibitem[]{} 
Miller, E. D., Bregman, J. N., \& Knezek, P. M. 2002, ApJ, 569, 134
\bibitem[]{} 
Myers, A. D., Outram, P. J., Shanks, T., Boyle, B. J., Croom, S. M., Loaring,
N. S., Miller, L., \& Smith, R. J. 2003, MNRAS, 342, 467
\bibitem[]{} 
Narayanan, A., Misawa, T., Charlton, J. C. \& Kim, T.-S. 2007, ApJ, 660, 1093 
\bibitem[]{} 
Navarro, J. F., Frenk, C. S., \& White, S. D. M. 1997, ApJ, 490, 493
\bibitem[]{} 
Nestor, D. B., Turnshek, D. A., \& Rao, S. M. 2005, ApJ, 628, 637 (NTR05)
\bibitem[]{} 
Nestor, D. B., Turnshek, D. A., \& Rao, S. M. 2006, ApJ, 643, 75 (NTR06)
\bibitem[]{} 
Nestor, D. B., Turnshek, D. A., Rao, S. M. \&  Quider, A. M.,  2007, ApJ, 658,
185
\bibitem[]{} 
Perlman, E. S.. Horner, D. J., Jones, L. R., Scharf, C. A., Ebeling, H.,
Wegner, G., \&  Malkan, M. 2002, ApJS, 140, 265
\bibitem[]{} 
Petitjean P., \& Bergeron J., 1990, A\&A, 231, 309
\bibitem[]{} 
Pisano, D. J., Barnes, D. G., Gibson, B. K., Staveley-Smith, L., Freeman, K.,
\& Kilborn, V. A., 2007, ApJ (arXiv:astro-ph/0703279)
\bibitem[]{} 
Prochaska, J. X., \& Herbert-Fort, S. 2004, PASP, 116, 622
\bibitem[]{}
Prochaska, J. X., Hennawi, J. F. \& Herbert-Fort, S. 2007,
arXiv:astro-ph/0703594 
\bibitem[]{} 
Prochter, G. E., Prochaska, J. X., \& Burles, S. M.  2006, ApJ, 639, 766
(PPB06) 
\bibitem[]{} 
Rao, S.M., \& Turnshek, D.A. 2000, ApJS, 130, 1
\bibitem[]{} 
Rao, S. M., Turnshek, D. A., \& Nestor, D. B.  2006, ApJ, 636, 610 
\bibitem[]{}
Rigby, J. R., Charlton, J. C., \& Churchill, C. W. 2002, ApJ, 565, 743
\bibitem[]{}
Russell, D. M., Ellison, S. L. \&  Benn, C. R. 2006, MNRAS, 367, 412
\bibitem[]{} 
Schneider, D. P. et al. 2005, AJ, 130, 367
\bibitem[]{} 
Scranton, R., et al. 2005, ApJ, 633, 589
\bibitem[]{} 
Smette, A, Claeskens J.-F.\& Surdej, J. 1997, New Astronomy, 2, 53
\bibitem[]{} 
Sparks, W. B., Carollo, C. M., \& Macchetto, F. 1997, ApJ, 486, 253
\bibitem[]{} 
Spergel, D. N., Bean, R., Dore , O.,
Nolta, M. R., Bennett, C. L., Hinshaw, G., Jarosik, N., Komatsu, E., Page, L.,
Peiris, L., Verde, L., Barnes, C., Halpern, M., Hill, R. S., Kogut, A., Limon,
M., Meyer, S. S., Odegard, N., Tucker, G. S., Weiland, J. L., Wollack, E., \&
Wright, E. L. 2007, arXiv:astro-ph/0603449
\bibitem[]{} 
Steidel, C. C., Kollmeier, J. A., Shapley, A. E., Churchill, C. W.,
Dickinson, M., \& Pettini, M., 2002, ApJ, 570, 526
\bibitem[]{} 
Steidel, C. C., \& Sargent, W. L. W. 1992, ApJS, 80, 1 
\bibitem[]{} 
Stocke, J. T., Morris, S. L., Gioia, I. M., Maccacaro, T.,
Schild, R., Wolter, A., Fleming, T. A., \& Henry, J. P. 1991, ApJS, 76, 813
\bibitem[]{} 
Takei, Y., Henry, J. P., Finoguenov, A., Mitsuda, K., Tamura, T.,
Fujimoto, R., \& Briel, U. G., 2007, ApJ, 655, 831    
\bibitem[]{} 
Tytler, D., Boksenberg, A., Sargent, W.L.W., Young, P., \& Kunth, D. 1987,
ApJS, 64, 667 
\bibitem[]{} 
Verheijen, M., van Gorkom, J., Szomoru, A., Dwarakanath, K. S.,
Poggianti, B., \& Schiminovich, D., 2007, NewAR, 51, 90
\bibitem[]{} 
Vollmer, B., Soida, M., Beck, R., Urbanik, M., Chyży, K. T.,
Otmianowska-Mazur, K., Kenney, J. D. P., \& van Gorkom, J. H. 2007, A\&A 464,
37 
\bibitem[]{} 
Vollmer, B., Braine, J. Combes, F., \& Sofue, Y. 2005, A\&A 441, 473 
\bibitem[]{} 
White S. D. M., Navarro J. F., Evrard A. E., \& Frenk C. S. 1993, Nature, 366,
429 
\bibitem[]{}
Williger, G. M., Campusano, L. E., Clowes, R. G. \&  Graham, M. J. 2002, ApJ,
578, 708 
\bibitem[]{}
Wolfe, A. M., Howk, J. C., Gawiser, E., Prochaska, J. X., \& Lopez, S. 2004,
ApJ, 615, 625 
\bibitem[Yee \& Ellingson(2003)]{yee2003} Yee, H.~K.~C., \& 
Ellingson, E.\ 2003, \apj, 585, 215 
\bibitem[]{}
York et al. 2000, AJ, 120, 1579
\bibitem[]{}
Zibetti, S. M\'enard, B., Nestor, D. B., Quider, A. M., Rao, S. M., \&
Turnshek, D. A. 2007, ApJ, 658, 161
\bibitem[]{}
Zwaan M. A., et al. 2003, AJ, 125, 2842
\end{thebibliography}
\end{document}